\documentclass[11pt]{article}
\usepackage{amsthm,amsfonts,amsmath}
\swapnumbers
\font\notefont=cmsl8

\theoremstyle{plain}

\newtheorem{thm}{THEOREM}[section]
\newtheorem{lm}[thm]{LEMMA}
\newtheorem{cl}[thm]{COROLLARY}

\theoremstyle{definition}

\theoremstyle{remark}
\newcommand{\R}{{\mathord{\mathbb R}}}
\newcommand{\RR}{{\mathord{\mathbb R}}^3}
\newcommand{\C}{{\mathord{\mathbb C}}}

\newcommand{\supp}{{\mathop{\rm supp\ }}}

\newcommand{\an}{{\mathord{a^{\phantom{*}}_\lambda}}}

\newcommand{\cre}{{\mathord{a^*_\lambda}}}

\begin{document}

\title{\bf{Ground States in Non-relativistic Quantum Electrodynamics}}
\author{\vspace{5pt} Marcel Griesemer$^1$, Elliott H. Lieb$^2$ and
Michael Loss$^{3}$ \\
\vspace{-4pt}\small{$1.$ Department of Mathematics, University of Alabama at 
Birmingham,} \\
\small{Birmingham, AL 35294}\\
\vspace{-4pt}\small{$2.$ Departments of Physics and Mathematics, Jadwin
Hall,} \\
\small{Princeton University, P.~O.~Box 708, Princeton, NJ
  08544}\\
\vspace{-4pt}\small{$3.$ School of Mathematics, Georgia Tech,
Atlanta, GA 30332}  \\ }
\date{\small September 18, 2000}
\maketitle

\footnotetext[1]{ Work partially
supported by the Faculty Development Program of UAB}
\footnotetext
[2]{Work partially
supported by U.S. National Science Foundation
grant PHY 98-20650-A01.}
\footnotetext
[3]{Work partially
supported by U.S. National Science Foundation
grant DMS 00-70589.\\
\copyright\, 2000 by the authors. This paper may be reproduced, in its
entirety, for non-commercial purposes.}

\begin{abstract}
The excited states of a charged particle interacting with the quantized
electromagnetic field and an external potential all decay, but such a
particle should have a true ground state ---  one that minimizes the energy
and satisfies the Schr\"odinger equation. We prove quite generally that this
state exists for {\it all values} of the fine-structure constant and
ultraviolet cutoff. We also show the same thing for a many-particle system
under physically natural conditions.  \end{abstract}

\section{INTRODUCTION}

An established picture of an atom or molecule is that even in the presence of a
quantized radiation field there is a ground state.  The excited states that
exist in the absence of coupling to the field are expected to melt into
resonances, which means that they eventually decay with time into the ground
state plus free photons.  This picture has been established by Bach, Fr\"ohlich
and Sigal in \cite{BFS2} for sufficiently small values of the various parameters
that define the theory.  Here we show that a ground state exists for all values
of the parameters (including a variable $g$-factor) in the one particle case
and, under a physically appropriate assumption, in the many-particle case.

We know that the Hamiltonian for the system is bounded below, so a ground
state energy always exists in the sense of being the infimum of the
spectrum, but the existence of a genuine normalizable solution to the
eigenvalue equation is a more delicate matter that has received a great deal
of attention, especially in recent years.  A physically simple example where
no ground state exists (as far as we believe now) is the free particle
(i.e., particle plus field).  In the presence of an external potential,
however, like the Coulomb potential of a nucleus, a ground state should
exist.

The difficulty in establishing this ground state comes from the fact that the
bottom of the spectrum lies in the continuum (i.e., essential spectrum), not
below it, as is the case for the usual Schr\"odinger equation. We denote the
bottom of spectrum of the free-particle Hamiltonian for $N$ particles with
appropriate statistics by $E^0(N)$. The `free-particle' Hamiltonian includes
the interparticle interaction (e.g., the Coulomb repulsion of electrons) but
it does not include the interaction with a fixed external potential, e.g.,
the interaction with nuclei. When the latter is included we denote the bottom
of the spectrum by $E^V(N)$.  It is not hard to see in many cases that
$E^V(N) < E^0(N)$, but despite this inequality $E^V(N)$ is, nevertheless, the
bottom of the essential spectrum.  The reason is that we can always add
arbitrarily many, arbitrarily `soft' photons that add arbitrarily little
energy.  It is the soft photon problem that is our primary concern here.

The main point of this paper is to show how to overcome this infrared problem
and to show, quite generally for a one-particle system, that a ground state
exists for all values of the particle mass, the coupling to the field
(fine-structure constant $\alpha = e^2/\hbar c$), the magnetic $g$-factor, and
the ultraviolet cutoff $\Lambda$ of the electromagnetic field frequencies,
provided a bound state exists when the field is turned off.
This result implies, in particular, that for a fixed ultraviolet cutoff
renormalization of the various physical quantities will not affect
the existence of a ground state. Of course, nothing
can be said about the  limit as the cutoff tends to infinity.
We also include a
large class of interactions much more general than the usual Coulomb
interaction.

The model we discuss has been used quite frequently in field theory.
In its classical version it was investigated by Kramers \cite{K} who seems to 
have been
the first to point out the possibility of renormalization. The quantized
version was investigated by Pauli and Fierz \cite{FP} in connection with
scattering theory. Most importantly, it was used by Bethe \cite{B} to obtain
a suprisingly good value for the Lamb shift. 

Various restricted versions of the problem have been attacked successfully.
In the early seventies Fr\"ohlich investigated the infrared problem in 
translation invariant models of scalar electrons coupled to scalar bosons 
\cite{JF1}.
In \cite{JF2} he  proved that for an electron coupled to a massive 
field a unique ground state exists for fixed total momentum. 
External potentials were not considered in these papers.

The first rigorous result on the bound state problem, to our knowledge, 
is due to Arai and concerns one
particle confined by an $x^2$-potential, the interaction with the photons
being subject to the dipole-approximation. For this model, which is
explicitly solvable, Arai proved existence and uniqueness of the ground
state \cite{AA}. Later, Spohn showed by perturbation theory that this result
extends to perturbed $x^2$- potentials \cite{HS0}. No bounds on the
parameters were needed to obtain these results but the methods oviously do
not admit extensions to more realistic models.

Bach , Fr\"ohlich and Sigal \cite{BFS0, BFS1} initiated the study of the full
nonrelativistic QED model (the same model as considered in the present paper)
under various simplifying assumptions. In \cite{BFS2} the  existence of a
ground state in this model for particles subject to an external binding
potential was proved for $\alpha\Lambda$ small enough. The main achievement
of this paper is the elimination of the earlier simplifying assumptions,
especially  the infrared regularization.  This is the first, and up to now
the only  paper where a `first principles' QED model was successfully
analyzed, but with a restricted parameter range.  A weaker result for the
same system but with simplifications such as infrared regularization were
independently obtained by Hiroshima by entirely different methods; he also
showed uniqueness of the ground state, assuming its existence \cite{FH}.

In a parallel development Arai and Hirokawa \cite{AH1,AH2}, Spohn \cite{HS2},
and Gerard \cite{CG} investigated the ground-state problem for systems
similar to the one of Bach et al in various degrees of mathematical
generality. Arai and Hirokawa proved existence in what they called a
`generalized spin-boson model'. If specialized to the case of a
non-relativistic $N$-particle system interacting with bosons, their result
proves the existence of a ground state for confined particles and small
$\alpha$. This result was extended in a recent preprint \cite{AH2} to account
for non-confined particles and systems with the true infrared singularity of
QED.  Concrete results in the infrared-singular case concern only special
models, however, such as the Wigner-Weisskopf Hamiltonian which describes a
two-level atom. Hirokawa continued this work in a recent preprint \cite{MH}.
Spohn and Gerard both proved existence of a ground state for a confined
particle and arbitrary coupling constant, the result of Gerard being somewhat
more general \cite{HS2,CG}.

For the existence of a ground state in the case of massive bosons, which is a
typical intermediate result in the cited works, a short and elegant proof was
given by Derezinski and Gerard \cite{DG, CG} for the case of linear coupling,
in which the $A^2$ term is omitted, and a confining potential. Some of
their ideas are used in our paper.

The Hamiltonian for $N$-particles has four parts which are described
precisely in the next section,
\begin{equation}
H^V= T +V +I +H_f \  .
\end{equation}

The dependence of $H^V$ on $N$ is not noted explicitly.  The first term,
$T=\sum_{j=1}^N T_j$, is the kinetic energy with `minimal coupling' in the
Coulomb gauge (i.e., $p$ is replaced by $p+\sqrt{\alpha}\, A$, where $A$ is the
magnetic vector potential satisfying $\mathrm{div} A=0$), $V(X)$, with $X= (x_1,
x_2, ..., x_N) $ is the external potential, typically a Coulomb attraction to
one or more nuclei.  In any case, we assume that $V$ is a sum of one-body
potentials, i.e.,
$$
V(X) = \sum_{j=1}^N v(x_j)\ .
$$

The  particle-particle interaction, $I$, has the important feature that it is
translation invariant and, of course, symmetric in the particle labels.  Both
$I$ and $V$ could be spin dependent, but we shall not burden the notation
with this latter possibility.  Typically $I$ is a Coulomb repulsion, but we
do not have to assume that $I$ is merely a sum of two-body potentials.  The
only requirements are: (1) the negative part of $v$  vanishes at infinity;
(2) the negative part of $I$ satisfies clustering, i.e., the intercluster part
of $I_-$ tends to 0 when the spacing between any two clusters tends to
infinity; (3) $V_- +I_-$ are dominated by the kinetic energy as in
(\ref{epsilonbound}). Another assumption we make (in the
$N$-particle case) is that there is binding, as described below.  

The natural choice for the kinetic energy is the `Pauli operator'
$T=(p+\sqrt{\alpha}\, A(x))^2 +\sqrt{\alpha}\, \sigma\cdot B$, but we can
generalize this to include the case of the usual kinetic energy
$T=(p+\sqrt{\alpha}\, A(x))^2$ by introducing a `g-factor', $g\in \R$.  Thus, we
take 
\begin{equation}\label{eq:kinetic} T=(p+\sqrt{\alpha}\, A(x))^2
+\frac{g}{2}\sqrt{\alpha}\, \sigma\cdot B \ .  
\end{equation} 
Note that $T$ is a
positive operator if $0\leq g \leq 2$ and may not be otherwise.  Nevertheless,
the Hamiltonian is always bounded below because of the ultraviolet cutoff we
shall impose on the $A$ field, which implies that $(g/2)\sqrt{\alpha}\,
\sigma\cdot B +H_f$ is always bounded below.

We believe that the `relativistic' operator $T= |p+\sqrt{\alpha}A(x)|$,
presents no real difficulty either, but we do not want to overburden this
paper with a lengthy proof. This problem is currently under investigation.

A model that is frequently discussed is the `Pauli-Fierz' model, but it is
not entirely clear how this is defined since several variants appear in
\cite{FP}. One version uses $T=(p+\sqrt{\alpha}A(x))^2 +\sqrt{\alpha}\,
\sigma\cdot B$, which is one of the models under consideration.  Another
variant uses a linearized version of this operator,
$T=p^2+2\sqrt{\alpha}p\cdot A(x) +\sqrt{\alpha}\, \sigma\cdot B$ or
$T=p^2+2\sqrt{\alpha}p\cdot A(x) $.  These variants are {\it not gauge
invariant} and, therefore, depend on the choice of gauge for $A$.  Our method
is applicable to these linearized models in some gauges, but not in others. 
We omit further discussion of this point since these variants are not the
most relevant ones  for quantum electrodynamics.

There is one important point that as
far as we know, has not been mentioned in the QED context. This is the binding
condition.  Our proof of the existence of a ground state uses, as input, the
assumption that 
\begin{equation}\label{eq:binding}
E^V(N) < E^V(N') + E^0(N-N')  \quad\quad {\rm for \ all} \ N'<N.
\end{equation}

Our work can be generalized (but we shall not do so here) to the case in
which the external potential is that of attractive nuclei and these nuclei
are also dynamical particles.  Then, of course, it is necessary to work in
the center of mass system and then (\ref{eq:binding}) must be replaced 
by the condition that $E^V(N) $ is less than the lowest two-cluster threshold.
While this condition, or (\ref{eq:binding}) in the static case, are
physically necessary for the existence of a ground state, the validity of
these conditions cannot be taken for granted. 

We prove  inequality (\ref{eq:binding}) for one particle ($N=1$) quite
generally, using only the assumption  that the ordinary Schr\"odinger
operator $p^2 +v$ has a negative energy ground state.  This certainly holds
for the Coulomb potential.  Indeed, one could expect, on physical grounds,
that there could be binding even if $p^2 +v$ is has no negative energy bound
state, because the interaction with the field increases the effective mass of
the particle --- and hence the binding energy. The same argument shows that
when there are $N$ particles at least one them is necessarily bound, i.e.,
$E^V(N) <  E^0(N)$.
When we consider more than one particle, we are not able to show
(\ref{eq:binding}) for all $N'>1$, even if $\sum p_j^2 +I+V$  has a ground
state. 

In the Coulomb case, it is possible to show (but we shall not do so here)
that condition (\ref{eq:binding}) is satisfied if the nuclear charge $Z$ is
large enough. The basic idea is that if breakup into two groups, one of them
with $N'$ particles close to the nucleus and the second consisting of $N-N'$
particles far away occurs, then there will be an attractive Coulomb tail
acting on the separated particles at a distance $R$ away with net attractive
potential $(Z-N')/R$. However, to localize one of these particles within
a distance $R$ of the nucleus will require a field energy localization error
of the order of $C/R$, by dimensional analysis arguments.  If $(Z-N') > C$
then the energy can be decreased by bringing one of the unbound particles
close to the nucleus.

Section \ref{sec:def} introduces the precise definition of our problem and
the main result Theorem \ref{t1}.

In Section 3, we show how to prove that $E^V(1)  < E^0(1)$. More generally,
$E^V(N)  <  E^0(N)$ if $v\leq 0$.

Our strategy to establish a ground state is the usual one of showing that a
minimizing sequence of trial vectors for the energy actually has a weak limit
that, in fact, is a minimizer. The problem here is that one can easily construct
minimizing sequences that converge weakly to zero by choosing vectors with
too many soft photons. To avoid this we take a special sequence.

To define this sequence we first consider an artificial model in which the
photons have a mass, i.e., $\omega(k) =\sqrt{k^2 +m^2}$. Here there is no
soft photon problem and we show in Section 4 that this model has a ground
state  $\Phi_m$.

In Section 5 we show that as $m\to 0$ the $\Phi_m$ sequence is minimizing.
Then in Section 6 we use the Schr\"odinger equation for $\Phi_m$
to deduce certain properties of $\Phi_m$ which we call infrared bounds.
One of these was proved in \cite{BFS2} but we need one more, which is new.

With these bounds we can show in Section 7 that $\Phi_m$ has a strong
limit as $m\to 0$, which is a minimizer for $H^V$.

Acknowledgment: We thank Professor Fumio Hiroshima for a useful correspondence
concerning equation (\ref{Hcommutator}) and for sending us his preprint 
\cite{FH2}.

\section{DEFINITIONS AND MAIN THEOREM}\label{sec:def}

The Hamiltonian for  N particles interacting with the quantized
radiation field and with a given external potential $V(X)$,
with $X = (x_1, x_2, ... ,x_N)$ and $x_j\in \R^3$,
is
\begin{equation}
H^V = \sum_{j=1}^N \left\{\left(p_j + \sqrt{\alpha} A (x_j)\right)^2 + \frac{g}{2}
\sqrt{\alpha}\,  \sigma_j \cdot B(x_j) \right\} + V(X) +I(X)
+ H_f.
\label{ham}
\end{equation}
The unit of energy is $Mc^2$/2, where $M$ is the particle mass, the
unit of length is $\ell_c = 2 \hbar/Mc$, twice the Compton wavelength
of the particle, and $\alpha = e^2/\hbar c$ is the dimensionless ``fine
structure constant'' (=1/137 in nature). The electric charge of the
particle is $e$. The unit of time is the time it takes a light wave to travel
a Compton wavelength, i. e., the speed of light is $c=1$.

The operator $p = -i \nabla$, while $A$ is the
(ultraviolet cutoff) magnetic vector potential (we use the Coulomb, or
radiation gauge). The unit of $A^2(x)$ is $Mc^2/2\ell_c$. The magnetic
field is $B$ = curl$A$ and the unit of B is $\alpha^{3/2}$ times the
quantity $M^2e^3c/4\hbar^{3}$, which is the value of $B$ for which
the magnetic length ($\hbar c/eB)^{1/2}$ equals the Bohr radius
$2\hbar^2/Me^2$.

The reader might wonder why we use these units, which seem to be more
appropriate for a relativistic theory than for the nonrelativistic theory we
are considering.  Why not use the Bohr radius as the unit of length, for
example? Our reason is that we want to isolate the electric charge, which is
the quantity that defines the interaction of matter with the electromagnetic
field, in precisely one place, namely $\alpha$. `Atomic units' have the
charge built into the length, etc. and we find this difficult to disentangle.

The Hilbert space is an appropriate subspace of 
$$ \mathcal{H} = \otimes^N L^2(\R^3;\C^2) \otimes
\mathcal{F}, 
$$ 
where $\mathcal{F}$ is the Fock space for the photon field.  We have in mind
Fermi statistics (the antisymmetric subspace of $\otimes^N L^2(\R^3;\C^2)$)
and the $\C^2$ is to accomodate the electron spin.  We can also deal with
"Boltzmann statistics",in which case we would set $g=0$ and use $\otimes^N
L^2(\R^3)$, or  bose statistics, in which case we would set $g=0$ and use the
symmetric subspace of $\otimes^N L^2(\R^3)$. These generalizations are
mathematically trivial and we do not discuss them further.

For our  purposes  we assume  that for every
$\varepsilon > 0$  there exists a constant $a(\varepsilon)$ such
that the negative part of the potentials, $V_-(X)$ and $I_-(X)$, satisfy
\begin{equation}\label{epsilonbound}
V_- +I_- \leq \varepsilon  \sum_{j=1}^N p_j^2 +
a(\varepsilon)
\end{equation}
as quadratic forms on ${\mathcal H}$.

The vector potential is
\begin{equation}
A(x) = \sum_{\lambda=1,2} \int_{|k|<\Lambda} ~ \frac{1}{\sqrt{|k|}}
\left[\varepsilon_{\lambda} (k) \an(k) + \varepsilon_{\lambda}
(-k)a^{\ast}_{\lambda}
(-k)\right] e^{ik \cdot x} d^3k \label{apot}
\end{equation}
where the operators $\an, a^{\ast}_{\lambda}$
satisfy the usual commutation relations
\begin{equation}
[\an(k), a^{\ast}_{\nu} (q)] = \delta ( k-q)
\delta_{\lambda, \nu}\ , ~~~ [\an(k), a_{\nu} (q)] =
0, \quad {\rm{etc}}
\end{equation}
and the vectors $\varepsilon_{\lambda}(k)$ are the two possible
orthonormal polarization vectors perpendicular to $k$. They are chosen for
convenience in (\ref{eq:pol1},\ref{eq:pol2}).

The number $\Lambda$ is the ultraviolet cutoff on the wavenumbers $k$. Our
results hold for all finite $\Lambda$. The details of the cutoff in
(\ref{apot})  are quite unimportant, except for the requirement that rotation
symmetry in $k$-space is maintained. E.g., a gaussian cutoff can be used
instead of our sharp cutoff. We avoid unnecessary generalisations.

The field energy $H_f$, sometimes called $d\Gamma(\omega)$ is given by
\begin{equation}\label{eq:fielden}
H_f = \sum_{\lambda=1,2} ~ \int_{\R^3} ~ \omega(k)
\cre(k) \an(k) d^3 k
\end{equation}

The energy of a photon is  $\omega(k)$ and the physical value of
interest to us is
\begin{equation}\label{eq:kay}
\omega (k) = |k|,
\end{equation}
in our units. Indeed, any continuous function that is bounded
below by ${\rm const.}|k|$ for small $|k|$ is acceptable. In the process of
proving the existence of a ground
state for $H$ we will first study the unphysical `massive photon'
case, in which
\begin{equation}\label{eq:kaym}
\omega_m(k) = \sqrt{k^2 + m^2}
\end{equation}
for some $m > 0,$ called the `photon mass'.

In the remainder of this paper, unless otherwise stated, we shall always
assume that there is no restriction on $\alpha,\ \Lambda$ and $g$ and that
$\omega(k)$ can be either as in (\ref{eq:kay}) or as in (\ref{eq:kaym}).

By Lemma \ref{lm:kinbound} we see easily that $H^V$ is bounded below for all
values of the parameters, including $m=0$. Thus, $H^V$ defines a closable
quadratic form and hence a selfadjoint operator, the Friedrich's extension.
We denote this extension again by $H^V$. 

Our main theorem is

\begin{thm}[Existence of a ground state]\label{t1}
Assume that the binding condition (\ref{eq:binding}) and the 
condition 
(\ref{epsilonbound}) hold. Then there is a vector $\Phi$ in the
$N$-particle Hilbert space $\mathcal{H}$ such that
\begin{equation}
H^V \Phi = E^V(N) \Phi  \   .
\end{equation}
\end{thm}

\section{UPPER BOUND}\label{sec:upper}
We shall prove the binding condition (\ref{eq:binding}) for one particle and,
with an additional assumption, for the $N$ particle case as well. This
is that if $N$-particles are present then at least one of them binds.

As we mentioned before, at least in the single particle case, that our
requirement that the system without the radiation field has a bound state is
somewhat unnaturally restrictive, since one expects that the radiation field
{\it enhances binding;} this has been shown to be true in the `dipole', or
Kramers approximation \cite{HS}.  We are able to show in the {\it one-particle}
case, that the photon field cannot decrease the binding energy.  It is quite
possible that there could be binding even when the operator $p^2+v$ does not
have a negative energy state, but we cannot shed any light on that question.

For the one-particle case the situation is less
delicate than the $N$-particle case. 

\begin{thm}[Binding of at least one particle] \label{thm:onebind}
Assume that the one-particle Hamiltonian $p^2 + v(x)$ has a negative energy
bound state with eigenfunction $\phi(x)$ and energy $-e_0$. Then, 
\begin{equation}
E^V(1) \leq E^0 (1)- e_0 \ ,
\end{equation}
i.e., binding continues to exist when the field is turned on. 

For the $N$-particle case we make the additional assumption that 
$v(x)\leq 0$ for all $x$. Then,
\begin{equation}
E^V(N) \leq E^0 (N)- e_0 \ ,
\end{equation}
i.e., at least one particle is bound.
\end{thm}

\begin{proof}
It suffices to prove that $E^V(N) \leq E^0(N) + \varepsilon -e_0$ for all
$\varepsilon >0$. There is a normalized vector $F \in {\cal H}$ such that
$\left(F, H^0 F \right) < E^0(N) +\varepsilon$. ($F$ is 
antisymmetric according to the Pauli principle.) We use the notation
$\langle \cdot, \cdot \rangle$ to denote the inner product in Fock space and spin space.
Then we  
can write $\left(F, H^0 F \right)=\int G(X) d^{3N}X$ with
\begin{eqnarray}
& & G(X) = \nonumber \\ 
& &\sum_{j=1}^N \left\{ \langle (-i\nabla_j + \sqrt \alpha A(x_j))F, 
(-i\nabla_j + \sqrt \alpha A(x_j))F \rangle (X) 
 + \sqrt{\alpha}\, (g/2) \langle F, \sigma_j\cdot B(x_j) F \rangle (X) 
\right\} \nonumber \\
& & + \langle F, (I + H_f) F\rangle (X)  \ . 
\end{eqnarray}

As a (unnormalized) variational trial vector we take the vector 
$\psi = \left[\sum_{j=1}^N\phi(x_j)^2\right]^{1/2}F$. Recall that 
$\phi(x) \geq 0$ since $\phi$ is the ground state of $p^2 +v$. We also recall
the Schwarz inequality 
\begin{equation}\label{schwarz}
\left|\frac{\sum_{j=1}^N \phi(x_j)\nabla \phi(x_j)}
{\left[\sum_{j=1}^N\phi(x_j)^2\right]^{1/2}}\right|^2
\leq \sum_{j=1}^N \left|\nabla \phi(x_j)\right|^2 \ .
\end{equation}

Using (\ref{schwarz}), integration by parts, and the fact that $\phi$ satisfies the 
Schr\"odinger equation
$(p^2+v)\phi=-e_0\phi$, we easily find that
\begin{eqnarray}\label{intF}
& &\left( \psi, \left[H^V-(E^0(N)+\varepsilon -e_0) \right] \psi \right)
\nonumber \\
& &\leq \int \left\{G(X) -(E^0(N) + \varepsilon )
\langle F,F\rangle (X) \right \} 
\sum_{j=1}^N  \phi(x_j)^2 \  d^{3N} X  \nonumber \\
& & +\int \sum_{j\not= k}  v(x_k)\phi(x_j)^2 \langle F,F\rangle (X)  
d^{3N} X  \  .
\end{eqnarray}

When $N=1$ the  last term in (\ref{intF}) is not present so no assumption about
the potential $v$ is needed. When $N>1$ we can omit the last term because it 
is negative by assumption.

Now, by the $\R^3$-translation invariance of $H_0$, for every $y\in \R^3$
there is a `translated' vector $F_y$ so that
$G(X) \to G(X+(y,...,y)) $ and $\langle F_y,F_y \rangle (X)  =
\langle F,F \rangle (X+(y,...,y))$. (This is accomplished by the unitary
operator on $\mathcal{H}$ that takes $x_j\to x_j+y$ for every $j$ and
$\an(k) \to exp(ik\cdot y) \an(k)$.)  Thus, if we denote the quantity in
$\{ \}$ in (\ref{intF}) by $W(X)$, and if we define $\psi_y$ by replacing
$F$ by $F_y$ in the definition of $\psi$, we have
\begin{eqnarray}
& & \Omega(y) =
\left(\psi_y, H^V - (E_0 +\varepsilon -e_0)\ \psi_y \right) \nonumber \\ 
& & \leq \int W(X+(y,...,y))\sum_{j=1}^N \phi(x_j)^2 d^{3N}X = 
\int W(X)\sum_{j=1}^N \phi(x_j-y)^2 d^{3N}X\ .
\end{eqnarray}

Note that $\int \Omega(y) dy \leq N \int  W(X)d^{3N}X$. But $\int W(X)d^{3N}X
=(F,(H^0 -E^0(N) -\varepsilon) \, F)$ and this is strictly negative
by assumption. Hence, for some $y\in \R^3$ we have that 
$\Omega(y) <0$ and thus $\psi_y \not= 0$, which proves the theorem.
\end{proof}

{\it Remark:} [Alternative theorem]

It may be useful to note, briefly, a different proof of Theorem
\ref{thm:onebind}, for {\it long range} potentials $v(x)$, such as the
attractive Coulomb potential $-Z/|x|$, which shows that the bottom of the spectrum of
$H^V$ lies strictly below $E^0$. Unfortunately, this proof does not show that
the difference is at least $e_0$. We sketch it for the one-body case.  Using
the notation of the proof above, the first step is to replace $F$ by
$F_R=u(x_1/R) F$ where $u$ is a smooth function with support in a ball of
radius $1$. One easily finds that $\left( F_R,H_0 \ F_R \right)/ \left( F_R,
\ F_R \right) =E^0 +\varepsilon +c/R^2$, where $c$ is a constant that depends
only on $u$ and not on $\varepsilon$ and $R$.  On the other hand $\left(
F_R,V \ F_R \right) / \left( F_R, \ F_R \right) \leq -Z/R$, to use the
Coulomb potential as an example. To complete the argument, choose $R=2c/Z$
and then choose $\varepsilon = c/R^2$.  What we have used here is the fact
that localization `costs' a kinetic energy $R^{-2}$, while the potential energy
falls off slower than this, e.g., $R^{-1}$.

\section{GROUND STATE WITH MASSIVE PHOTONS}\label{sec:mass}

As we emphasized in the introduction, not every minimizing sequence converges
to the minimizer for our $m=0$ problem, i.e., with $\omega(k)=|k|$. The
situation is much easier for the massive case (\ref{eq:kaym}).  The
Hamiltonian in this case is given by (\ref{ham}) and $H_f$ is given by
(\ref{eq:fielden}) with (\ref{eq:kaym}).  To emphasize the dependence on $m$
we denote this Hamiltonian and field energy by $H^V_m$ and $H_f(m)$,
respectively.  Likewise, $E^V(m,N)$ and $E^0(m,N)$  denote the mass dependent
energies, as defined before.  

We emphasize that the vector potential is still given by
(\ref{apot}), but we could, if we wished, easily replace $|k|^{-1/2}$ in
(\ref{apot}) by $(k^2+m^2)^{-1/4}$.  

It will be shown in this section that $H^V_m$ has a ground state.
More precisely we prove
\begin{thm}[Existence of ground state]\label{existence_with_mass}
Assume that for some fixed value of the ultraviolet cutoff $\Lambda$ there
is binding for the Hamiltonian $H^V_m$, i.e., $E^V(m,N) <
\Sigma^V(m,N)$ where $\Sigma^V(m,N)= min\{ E^V(m,N') + E^0(m,N-N'): {\rm
all}\ N'<N \}$ is the `lowest two-cluster threshold'.  Then $E(m,N)$ is an
eigenvalue, i.e., there exists a state $\Phi_m$ in ${\cal H}$ such that $H^V_m
\Phi_m = E(m,N) \Phi_m$.
\end{thm}
\begin{proof}
Let us first show
that it suffices to prove that for any normalized sequence
$\Psi^j$, $j=1,2,...$, (not necessarily minimizing) tending weakly to zero
\begin{equation}
\liminf_{j \to \infty}(\Psi^j, H^V_m \Psi^j) > E^V(m,N) \ .\label{weakzero}
\end{equation}
To prove this let $\Phi^j$ be some {\it minimizing} sequence, i.e., assume that
\begin{equation}
\Vert \Phi^j \Vert = 1\ ,
\end{equation}
and that
\begin{equation}
(\Phi^j, H^V_m \Phi^j) \to E^V(m,N) \ . \label{minsequence}
\end{equation}
By the Banach Alaoglu Theorem we can assume that this sequence,
as well as the sequence $H^V_m \Phi^j$ converge weakly in the sense that
for any $\Psi \in {\cal H}$ with $(\Psi, H^V_m \Psi) < \infty$
we have that
\begin{equation}
(\Psi, H^V_m \Phi^j) \to (\Psi, H^V_m \Phi_m) \ ,
\end{equation}
where $\Phi_m$ is the weak limit of $\Phi^j$. Our goal is to show that
$(\Phi_m,H^V_m \Phi_m) = E^V(m,N)$ and that  $\Vert \Phi_m \Vert =1$.

Write $\Phi^j = \Phi_m + \Psi^j$. Obviously $\Psi^j$ as well as
$H^V_m \Psi^j$ go weakly to zero.
Thus 
\begin{align*}
0 &= \lim_{j\rightarrow\infty}(\Phi^{j},(H^V_m-E^V(m,N))\Phi^{j}) \\
  &= \lim_{j\rightarrow\infty}((\Phi_m+\Psi^j),(H^V_m-E^V(m,N))(\Phi_m+\Psi^j)) \\
&=
  \lim_{j\rightarrow\infty}(\Psi^j,(H^V_m-E^V(m,N))\Psi^j)+(\Phi_m,(H^V_m-E^V(m,N))\Phi_m)
\end{align*}
where we used that the cross terms vanish. Since $H^V_m-E^V(m,N)\geq 0$
this shows that $\Phi_m$ minimizes the energy, and, furthermore, that
\begin{equation*}
  0\geq \lim_{j\rightarrow\infty}(\Psi^j,(H^V_m-E^V(m,N))\Psi^j)
  \geq \delta
  \liminf_{j\rightarrow\infty}\Vert\Psi^j\Vert^2
\end{equation*}
for some positive constant $\delta$. The second inequality is
trivial if $\liminf_{j\rightarrow\infty}\Vert\Psi^j\Vert^2=0$ and
otherwise follows from our assumption (\ref{weakzero}). This proves
that $\Psi^j$ converges strongly to zero along a subsequence, which implies
that $\Vert\Phi_m\Vert=1$. Hence $\Phi_m$ is a normalized
ground state. Thus, it suffices to prove (\ref{weakzero}).

The steps that lead to a proof of (\ref{weakzero}) are quite standard.
The only difficulty is that one has to localize in Fock space, which we 
describe first. We follow \cite{DG} with some necessary modifications and 
some simplifications. 

Recall that, when the $a_\lambda^\#$ operators are viewed in 
$x$-space
\begin{equation}
\an(f): {\cal F} \to {\cal F}\ ,\  \cre(g) :{\cal F} \to {\cal F}\ ,
\end{equation}
they obey the commutation relations
\begin{equation}
\left[ \an(f) , \cre(g) \right] = \int_{\R^3} \overline{f}(x) g(x) d^3 x
=:(f,g)\ .
\end{equation}

Consider now two  smooth localization functions $j_1$ and $j_2$ that satisfy
$j_1^2 + j_2^2 =1$ and $j_1$ is supported in a ball of radius $P$. The
first derivatives of $j_1$ and $j_2$ are of order $1/P$. 

The operators
\begin{equation}
c_{\lambda}(f) =\an(j_1f)\otimes {\cal I} + {\cal I} \otimes \an(j_2f)\ ,
c^*_{\lambda}(g) = \cre(j_1 g) \otimes {\cal I} + {\cal I} \otimes \cre (j_2 g)
\end{equation}
act both on the space ${\cal F} \otimes {\cal F}$.
Note that
\begin{equation}
\left[ c_{\lambda}(f),c^*_{\lambda}(g) \right] =(f,g) \ .
\end{equation}
Thus, these new creation and anihilation operators create another Fock
space ${\cal F}^l$ that is a subspace of ${\cal F} \otimes {\cal F}$
and is isomorphic to the old Fock space ${\cal F}$. Hence, there exists
a map
\begin{equation}
U:{\cal F} \to {\cal F}^l
\end{equation}
that is an invertible isometry between Fock spaces.
It is uniquely specified by the properties
\begin{equation}
a^{\#} = U^* c^{\#} U \ ,
\end{equation}
and the vacuum in ${\cal F}$ is mapped to the vacuum in ${\cal F} \otimes 
{\cal F}$.

The map $U^*$ is defined on ${\cal F}^l$ only, but we can extend it to
all of ${\cal F} \otimes {\cal F}$ by setting $U^*F=0$ whenever $F \in
{\cal F} \otimes {\cal F}$ is perpendicular to ${\cal F}^l$. In other 
words $U^*$ is a partial isometry between Fockspaces where $U^*U ={\cal I}$
on ${\cal F}$, and where $UU^*$ is the orthogonal projection onto ${\cal F}^l$.
We continue to denote the extended map by $U^*$.

Let $\phi$ and $\overline{\phi}$ be  smooth nonnegative functions,
with $\phi^2 + \overline{\phi}^2 =1$, $\phi$  identically one on the
unit ball,  and  vanishing outside the ball of radius $2$.
Set $\phi_R(X) = \phi(X/R)$.
It is a standard calculation to show that for any $\Psi$ with finite
energy
\begin{equation}
(\Psi, H^V_m \Psi) = (\phi_R \Psi, H^V_m \phi_R \Psi)
+(\overline{\phi_R} \Psi, H^V_m \overline{\phi_R} \Psi)
-(\Psi, (\nabla \phi_R)^2 \Psi) - (\Psi, (\nabla \overline{\phi_R} )^2 
\Psi)\label{IMS} \ .
\end{equation}
The last two terms in (\ref{IMS}) are bounded by $const./R^2$.

One goal will be to show that for any $\Psi$ with finite energy
\begin{eqnarray}
& & (\Psi,\phi_R H^V_m \phi_R \Psi)=\nonumber \\
& & (\Psi,\phi_RU^*\left\{H^V_m  \otimes {\cal I}
+ {\cal I} \otimes H_f \right\} U\phi_R \Psi) + o(1) \ . \label{locham}
\end{eqnarray}
The error term $o(1)$ vanishes as both $R$ and $P$ go to infinity
and depends otherwise only on the energy of $\Psi$. Notice that the
invertible map
$U$ depends on the cutoff parameter $P$ as well. (\ref{locham}) will be proved
in Lemma \ref{locallemma} in the Appendix.
The intuition behind the estimate (\ref{locham}) is that localized
electrons interact only weakly with far away photons. Those photons
are described solely by their own field energy.

An immediate consequence of (\ref{locham}) is the estimate
\begin{equation}
(\Psi,\phi_R H^V_m \phi_R \Psi) \geq (E^V(m,N)+m) \Vert \phi_R \Psi
\Vert ^2 -
m(\phi_R \Psi,U^*{\cal I} \otimes P_2U \phi_R \Psi) + o(1) \ , 
\label{inner}
\end{equation}
which is obtained by noting that the field energy in the second factor
can be estimated from below by
\begin{equation}
H_f \geq m{\cal I} - mP_2 \ ,
\end{equation}
where $P_2$ is the projection onto the vacuum of the second factor
of ${\cal F}\otimes {\cal F}$.

In a further step we prove in Lemma \ref{compactlemma} that the sequence
\begin{equation}
(\phi_R\Psi^j,U^* {\cal I}\otimes P_2 U\phi_R \Psi^j) \to 0 \label{compact}
\end{equation}
as $j \to \infty$. 

Returning to (\ref{IMS}), using Corollary \ref{threshold} we have that
\begin{equation}
(\overline{\phi_R} \Psi, H^V_m \overline{\phi_R} \Psi) \geq  
\Sigma^V(m,N) \Vert \overline{\phi_R} \Psi \Vert^2
 - o(1)
 \ , \label{electronloc}
\end{equation}
with $o(1)$ going to zero as $R \to \infty$. Roughly speaking 
$\overline{\phi}$ forces some
of the particles to be far away from the origin. 
Any such particle configuration can be described by two clusters with
no interaction between them. In particular the interaction between
these clusters via the radiation field is turned off. This means that
each cluster carries its own field energy.
To prove this the localization in Fock space is used.
Moreover the cluster  that is far away from 
the origin does not interact with the external potential although the
repulsion among its particles is still present.  

To summarize, by combining (\ref{inner}), (\ref{compact}) and 
(\ref{electronloc}) we have
proved that
\begin{equation}
\liminf(\Psi^j,H^V_m \Psi^j)
 \geq (E^V(m,N)+\delta)  + o(1)
\end{equation}
where
\begin{equation}
\delta = \min \{m, \Sigma^V(m,N) - E^V(m,N) \} \ ,
\end{equation}
and $o(1)$ tends to zero as $R \to \infty$ and $P \to \infty$.
\end{proof}

\section{A MINIMIZING SEQUENCE}\label{sec:minseq}

We consider the Hamiltonian $H^V_m$  defined in
(\ref{ham}) with field energy $H_f(m)$ defined using
$\omega_m(k)=\sqrt{k^2+m^2}$. Our main goal here is Theorem
\ref{minseq}, which shows that the ground states of the $m>0$ problem
form a minimizing sequence for the $m=0$ problem.

\begin{thm}[$E^V(m,N)$ converges to $E^V(0,N)$]\label{conv1}
As  $m \to 0$, 
\begin{equation}\label{conv2}
E^V(m,N) \to E^V(0,N) \ \ \ \ \mathrm{and}  \ \ \ \ E^0(m,N) \to E^0(0,N)
\end{equation}
\end{thm}

\begin{proof} First, note that $H^V_m > H^V_{m'} > H^V_0 $ if $m > m'>0$,
because $\omega_m$ has this same monotonicity property. Therefore, for any 
sequence of $m\to 0$, $E^V(m,N)$ is monotonically decreasing and has a
sequence-independent, finite limit, which we call $E^*$, and we note that
$E^* \geq E^V(0,N)$. To prove the opposite, namely $E^* \leq E^V(0,N)$,  we
shall prove that $E^* \leq E^V(0,N)+2\varepsilon$ for every $\varepsilon >
0$.

Let $\Phi \in \mathcal{H}$ be normalized and such that $\left(\Phi, H^V_0 \
\Phi\right) < E^V(0,N) +\varepsilon$. 

We note that $H^V_m < H^V_0 +m {\mathcal N}$, where ${\mathcal N}$ is the
number operator

\begin{equation}\label{numop}
{\mathcal N} = \sum_{\lambda=1,2} \ \int_{|k|<\Lambda} \ \cre(k) \an(k) d^3k.
\end{equation}

Thus, if we use $\Phi$ as a variational function for $H^V_m$ we have
$E^V(m,N) < E^V(0,N)
 +\varepsilon + m \left(\Phi, {\mathcal N} \ \Phi\right)$, and our goal
is accomplished {\it provided that } $\Phi$ can be chosen so that
$\left(\Phi, {\mathcal N} \ \Phi\right) <\infty$, in addition to $\left(\Phi,
H^V_0 \ \Phi\right) < E^V(0,N) +\varepsilon$. If a way can be found to
modify $\Phi$  to another vector $\widetilde\Phi$ so that 
$\left(\widetilde\Phi, {\mathcal N} \ \widetilde\Phi\right) <\infty$, 
in addition to $\left(\widetilde\Phi,
H^V_0 \ \widetilde\Phi\right) < E^V(0,N) +2\varepsilon$ the proof will 
be complete.

A suitable choice is $\widetilde\Phi = \Pi_n\Phi$, with $\Pi_n$ being the
projector onto the subspace of $\mathcal{H}$, with $n$ or fewer photons,
i.e., $\mathcal{H}_n = \otimes^N L^2(\R^3;\C^2) \otimes \mathcal{F}_{\leq n}$,
for an appropriately large $n$. It is easy to see, with the help of Lemma
\ref{lm:kinbound}  that $( \Pi_n\Phi,[V+I] \Pi_n\Phi) \to ( \Phi,[V+I]
\Phi)$, that $( \Pi_n\Phi,H_f \Pi_n\Phi) \to (\Phi,H_f \Phi)$, and $(
\Pi_n\Phi, p_j^2 \Pi_n\Phi) \to (\Phi,p_j^2 \Phi)$ as $n \to \infty$.  The
following Lemma \ref{photonconv}, shows that the other terms converge as
well. The same proof works for $E^0(m,N)$.  \end{proof}

\begin{lm}[Finite photon number approximation]\label{photonconv}
Let $\Phi \in \mathcal{H}$ be such that $\left(\Phi,H_f \Phi\right)
<\infty$.
Denote by $\Pi_n$ the
projection in $\mathcal{H}$ onto states with photon number less than or
equal to $n$, i.e., $\Pi_n$ is the projection onto the subspace
$\mathcal{H}_n = \otimes^N L^2(\R^3;\C^2) \otimes
\mathcal{F}_{\leq n} \subset \mathcal{H}$. Let $\Phi_n = 
\Pi_n \Phi$. Then  we have the following strong convergence as $n\to \infty$
(in addition to $\Phi_n  \to \Phi$ )
\begin{eqnarray}
A(x_j) \Phi_n  &\to & A(x_j) \Phi \nonumber \\
B(x_j) \Phi_n  &\to &   B(x_j) \Phi \ .
\end{eqnarray}
\end{lm}

\begin{proof}   
Write $A=D+D^*$ where $D$ contains the annihilation operators and $D^*$
contains the creation operators. We omit $(x_j)$ and we omit the vector index
of $A$ for simplicity. Since  $\left(\Phi,H_f \Phi\right)
<\infty$ we learn from Lemma \ref{lm:Abound}  that
$D\Phi$ and $D^*\Phi$ are in $\mathcal{H}$.
Since $\Pi_{n-1} \to \mathcal{I}$ strongly,
$$
 D\Phi_n = \Pi_{n-1}D\Phi \to D\Phi \ .
$$
The same  holds if $D$ is replaced by $D^*$ and $n-1$ by $n+1$. This proves
the statement for $A$. The statement for $B$ is proved in the same way.
\end{proof}

The following are two corollaries of Theorem \ref{conv1}. The first is the
input for Section \ref{sec:infrared}. The second is important for showing that
it is only necessary to state the `binding' condition (\ref{eq:binding})
once (for $m=0$). It is a trivial consequence of Theorem \ref{conv1}.

\begin{thm}[Minimizing sequence]\label{minseq}
Suppose that $m_1 > m_2 > ... >0$ is a sequence tending to zero and
suppose that $\Phi_j$, for $j=1,2,...$ is an approximate minimizer for
$H^V_{m_j}$ in the sense that
\begin{equation}
\delta_j \equiv \left( \Phi_j,  H^V_{m_j} \ \Phi_j \right) -E^V(m_j,N) \to 0
\ \ \ \  {\rm as} \  j \to \infty \ .
\end{equation}
Then $\Phi_1, \Phi_2, ... $ is a minimizing sequence for $H^V_0$.
\end{thm}

\begin{proof}
\ \ \ \ $E^V(m_j,N) +\delta_j  = \left( \Phi_j,  H^V_{m_j} \ \Phi_j \right)
> \left( \Phi_j,  H^V_0 \ \Phi_j \right) \geq E^V(0,N)  .  $

\end{proof}

\begin{lm}[Binding without mass implies uniform binding with mass]
\label{binding_with_mass}
Assume  (\ref{eq:binding}), i.e., assume that
$$
E^V(0,N) < \min\{E^V(0,N') + E^0(0,N-N'): N' <N\} -2 \varepsilon \ .
$$
Then, for all sufficiently small $m$,
$$
E^V(m,N) < \min\{E^V(m,N') + E^0(m,N-N'): N' <N\} - \varepsilon \ .
$$
\end{lm}

\begin{proof} Each of the various energies converges as $m\to 0$ by  
Theorem \ref{conv1}. 
\end{proof}

\section{TWO INFRARED  BOUNDS}\label{sec:infrared}

We have seen that any sequence of (approximate) minimizers for the $H^V_m$
problem ($m>0$) is a minimizing sequence for the $H^V_0$ problem. A crucial
point was the possibility of finding an approximate minimizer for the $H^V_0$
problem that has a finite expectation value for the total photon number
operator ${\mathcal N}$.

We also know by Corollary \ref{binding_with_mass} and Theorem
\ref{existence_with_mass} that the $H^V_m$ problem has a ground state,
$\Phi_m$, and in this section we shall prove two theorems about the soft
photon behavior of $\Phi_m$. The first, Theorem \ref{finitephot}, is about
the photon number $a_\lambda^*(k)a_\lambda^{\phantom{*}}(k)$ and is  based on
a  method of \cite{BFS2}.  The second, Theorem \ref{finitederiv}, which has no
antecedent we are aware of, is about the derivative of
$a_\lambda^*(k)a_\lambda^{\phantom{*}}(k)$  with respect to $k$.  

\begin{thm}[Photon number bound]\label{finitephot}
Assume that there is binding, i.e., $E^V(m,N) < \Sigma^V(m,N)$.
Assume that $\Phi_m$ is a normalized ground state for the many-body
Hamiltonian
$H^V_m $, $m\geq 0$. Then 
$$
\left( \Phi_m, \cre(k) \an(k)\Phi_m \right) < \frac{P \alpha}{|k|}(1+g^2)
\chi_{\Lambda}(k),
$$
where $P$ is a finite constant independent of $\Phi_m$, $g$,
$\alpha$, and depends on $m$ only via
the binding energy $\Sigma^V(m,N) - E^V(m,N) > 0$ and of course
on $\Lambda$ . The function
$\chi_{\Lambda}(k)$ is the characteristic function 
of the ball of radius $\Lambda$.
\end{thm}

{\it Remark:} We have proved in Corollary \ref{binding_with_mass} that the
binding energy is a uniformly positive function of $m$ for small $m$, if the
binding energy at $m=0$ is  not zero. Therefore, Theorem \ref{finitephot}
implies that the number operator $\mathcal{N}$ in (\ref{numop}) is uniformly
bounded for small $m$.

The proof of Theorem \ref{finitephot} will be based on the following  lemma
about exponential decay of eigenfunctions, concerning which there is a vast
literature (see \cite{AO, CT, RS, SA}). We do not strive at all to get the
best exponential decay constants. For us the only really relevant goal is a
decay estimate that depends {\it only} on $\Sigma^V(m,N) - E^V(m,N) > 0$, 
but not otherwise on $m$.

\begin{lm}[Exponential decay]\label{exp}
Let $H^V_m$ be the N-body Hamiltonian in (\ref{ham}) and let $\Phi_m$ be a
groundstate wave function, which necessarily satisfies the Schr\"odinger
equation
\begin{equation}\label{sch}
     H^V_m \Phi_m = E^V(m,N)\Phi_m \ .
\end{equation}
We assume that $\Sigma^V(m,N)-E^V(m,N) >0$ and choose $\beta>0$ with
$\beta^2<\Sigma^V(m,N)-E^V(m,N)$. Then 
\begin{equation}\label{norm}
\Vert \exp(\beta |X|) \Phi_m \Vert^2 \leq
C(1+\frac{1}{\Sigma^V(m,N)-E^V(m,N)-\beta^2})\Vert\Phi_m\Vert^2
\end{equation}
where the constant $C$ does not depend on $m$. 
\end{lm}

The strategy of the following proof is probably due to Agmon
\cite{SA}. We learned it from \cite{HuSi}.

\begin{proof}  
Let $G(X)$ any smooth, bounded function on $\R^{3N}$  with
bounded first derivative.
We easily compute
\begin{equation}
\left[ \ \left[H^V_m - E^V(m,N ) ,\  G\right], \ G \right] = -2|\nabla G|^2.
\end{equation}

We use the  Schr\"odinger equation (\ref{sch}) to compute
\begin{equation}\label{comm}
\left(G\Phi_m, [H^V_m -E^V(m,N)] G\Phi_m \right)  
= -\frac{1}{2} \left(\Phi_m, \left[ \ \left[H^V_m - E(m ) ,\ G\right], 
\ G \right] \Phi_m \right) = \left(\Phi_m, |\nabla G|^2 \Phi_m \right) \ .
\end{equation}

Now we choose $G$ to be
\begin{eqnarray}
G(X) &=& \chi (X/R) \exp [f(X)], \ \ \ {\rm where} \nonumber \\
 f(X) &=& \left[\frac{\beta |X|}{1+\varepsilon
|X|}\right] ,
\end{eqnarray}
and where $0\leq \chi \leq 1$ is a smooth cutoff function that is
identically equal to
1 outside the ball of radius 2, and identically zero inside the ball of
radius 1. We let $\varepsilon \to 0$ at the end.

Next, we calculate
$$
|\nabla G|^2 = |\nabla \chi|^2 e^{2f}
+2 \nabla \chi \cdot \nabla f e^f G + |\nabla f|^2 G^2,
$$
and note that the first and second terms are compactly supported
in $\R^{3N}$ and each is bounded by a constant $C$ that depends
on $\beta$ and $R$.

Returning to (\ref{comm}),  we obtain, after rearranging terms,
\begin{equation}\label{edbound1}
\left(G\Phi_m, \left( H^V_m -E^V(m,N) - |\nabla f|^2) \right) G\Phi_m
\right) \leq C \Vert \Phi_m \Vert ^2.
\end{equation}
Since $|\nabla f|\leq\beta$ we know by Corollary \ref{threshold} that
\begin{eqnarray}
& & \chi\left( H^V_m -E^V(m,N) -
|\nabla f|^2 \right)\chi \nonumber \\
 &\geq &\left( \Sigma^V(m,N) -E^V(m,N) -
\beta^2-o(1) \right)\chi^2\hspace{3em} (R\to\infty) \ .
   \end{eqnarray}
In conjunction with (\ref{edbound1}) this shows that \[ \Vert G\Phi_m\Vert^2
\leq \frac{2C}{\Sigma^V(m,N)-E^V(m,N)-\beta^2}\Vert\Phi_m\Vert^2\] for $R$
large enough. After letting $\varepsilon \to 0$ by monotone convergence a
similar bound with $G$ replaced by $\chi \exp(\beta |X|)$ is obtained. 
\end{proof}

\begin{proof}[ Proof of Theorem \ref{finitephot}] This proof is a slight
modification of the one in \cite{BFS2}. The basic idea is to show that there
is effectively no interaction between {\it localized } particles and low
momentum photons. To make this idea explicit we write our Hamiltonian in a
gauge different from the usual Coulomb gauge.

To be precise, define
\begin{equation}
\widetilde{A}(x) = A(x) -A(0),
\end{equation}
which is well defined owing to the ultraviolet cutoff.
The unitary operator that accomplishes this is
$U= \exp[i\sum_{j=1}^N \sqrt{\alpha}x_j \cdot A(0)]$. This is an `operator-valued
gauge transformation'. It commutes with $A(x)$ for all $x$, but not with
$\an(k)$ or with $H_f$.

Define 
\begin{equation}
b_\lambda(k,X) = U\an(k) U^* = \an(k) - i w_\lambda(k,X), \label{beeop}
\end{equation}
with $w_\lambda(k,X) = \chi_{\Lambda}(k)|k|^{-1/2}\varepsilon_\lambda(k)\cdot \sum_{j=1}^N
x_j$. The transformed Hamiltonian $\widetilde{H}_m$ is
\begin{gather}
\widetilde{H}_m = UH^V_m U^* =  \sum_{j=1}^N \left\{\left(p_j +\sqrt{\alpha}
\widetilde{A}(x_j) \right)^2 +\frac{g}{2}\sqrt{\alpha}\, \sigma_j 
\cdot B(x_j) \right\}+V +\widetilde{H}_f(m)\\
\widetilde{H}_f(m) = \sum_{\lambda = 1,2} \int_\RR \omega_m(k) b_\lambda^*(k,X) \
b_\lambda(k,X)\ .
\end{gather}

To estimate $\Vert\an(k)\Phi_m\Vert$,  write
\begin{equation}
\an(k) \Phi_m = U^* \an(k)\widetilde{\Phi}_m - i
w_\lambda(k,X)\Phi_m \label{relation}
\end{equation}
where $\widetilde{\Phi}_m=U\Phi_m$, and note that
\begin{equation}\label{estimate0}
\|w_\lambda(k,X)\Phi_m \| \leq \sqrt{N}
\frac{\chi_{\Lambda}(k)}{|k|^{1/2}}\| |X|\Phi_m \|.
\end{equation}
It remains to estimate $\|\an(k)\widetilde{\Phi}_m \|$. By
the Schr\"odinger equation for $\widetilde{\Phi}_m $
\begin{eqnarray}
\left( \widetilde{H}_m-E^V(m,N) \right) \an(k) \widetilde{\Phi}_m
&=&\left[ \widetilde{H}_m, \an(k) \right]\widetilde{\Phi}_m  \nonumber \\
&=&2 \sqrt{\alpha} |k|^{-1/2}\varepsilon_\lambda(k)\cdot \sum_{j=1}^N
(p_j + \sqrt{\alpha}\, \widetilde{A}(x_j))(1-e^{-ik\cdot x_j})
\widetilde{\Phi}_m  \nonumber \\
& & +i\frac{g}{2}\sqrt{\alpha}\, \frac{k\wedge \varepsilon_\lambda(k)}
{\sqrt{|k|}} \cdot \sum_{j=1}^N \sigma_j e^{-ik\cdot x_j}\widetilde{\Phi}_m 
 \nonumber \\ 
& & -\omega_m(k) b_\lambda(k,X) \widetilde{\Phi}_m  \ . \label{Hcommutator}
\end{eqnarray}

While this equation is correct, the derivation is somewhat formal. This
is rigorously justified in Appendix \ref{sec:appendixb}.

Now add $\omega_m(k) \an(k)\widetilde{\Phi}_m$ on both sides.
Since $E^V(m,N)$ is the ground state energy and $\omega_m(k) >0$ the operator
$\widetilde{H}_m-E^V(m,N)+\omega_m(k)$ has a bounded inverse $R(\omega_m(k))$ and hence
\begin{equation}\label{anequation}
\begin{split}
\an(k) \widetilde{\Phi}_m
= & 2\sqrt{\alpha} R(\omega_m(k))\frac{\chi_{\Lambda}(k)}{|k|^{1/2}}
\varepsilon_\lambda(k) \cdot \sum_{j=1}^N
(p_j + \sqrt{\alpha}\, \widetilde{A}(x_j)) (1-e^{-ik\cdot x_j})
\widetilde{\Phi}_m  \\
&  +i R(\omega_m(k)) {\chi_{\Lambda}(k)} \frac{g}{2}\sqrt{\alpha}\, \frac{k\wedge 
\varepsilon_\lambda(k)}
{\sqrt{|k|}} \cdot \sum_{j=1}^N \sigma_j e^{-ik\cdot x_j}\widetilde{\Phi}_m 
  \\ 
 & - R(\omega_m(k))\omega_m(k)iw_{\lambda}(k,X)\widetilde{\Phi}_m.
\end{split}
\end{equation}
For consistency, note that for $|k| > \Lambda$ $w_{\lambda}(k, X)=0$ and
 hence 
$\an(k)\widetilde{\Phi}_m = 0$, i.e, for these modes 
$\widetilde{\Phi}_m$ is the vacuum as it should be for
a minimizer.
Since $\|R(\omega_m(k))\|\leq \omega_m(k)^{-1}$ the norm of the last term is
bounded by
\begin{equation}
\sqrt{\alpha}\sqrt{N} \frac{\chi_{\Lambda}(k)}{|k|^{1/2}} \| |X|
\widetilde{\Phi}_m\|.
\end{equation}
To bound the norm of the first term we need to estimate
\begin{eqnarray}
& &\Vert \sum_{j=1}^N R(\omega_m(k))
\varepsilon_\lambda(k) \cdot (p_j +\sqrt{\alpha}\, \widetilde{A}(x_j)) (1-e^{-ik
\cdot x_j}) \widetilde{\Phi}_m \Vert \nonumber \\ &=&\sup_{\Vert
\eta \Vert \leq 1} \left|\sum_{j=1}^N \left(\varepsilon_\lambda(k) \cdot 
(p_j + \sqrt{\alpha}\,\widetilde{A}(x_j)) R(\omega_m(k)) \eta, (1-e^{-ik\cdot x_j})
\widetilde{\Phi}_m \right)\right| \nonumber \\ &\leq &\sup_{\Vert
\eta \Vert \leq 1} \left[\sum_{j=1}^N \Vert (p_j +
\sqrt{\alpha}\, \widetilde{A}(x_j)) R(\omega_m(k)) \eta \Vert ^2 \right]^{1/2} \left[
\sum_{j=1}^N \Vert (1-e^{-ik\cdot x_j})\widetilde{\Phi}_m \Vert^2
\right]^{1/2} \label{estimate1}
\end{eqnarray}
Next estimate the square of the first factor to get
\begin{eqnarray}
& &\left(\eta, R(\omega_m(k)) \left[ \sum_{j=1}^N (p_j + 
\sqrt{\alpha}\,\widetilde{A}(x_j))^2 \right] 
R(\omega_m(k)) \eta
\right) \nonumber \\
&\leq & a\left(\eta, R(\omega_m(k)) H^V_m R(\omega_m(k)) \eta  \right) + b \nonumber \\
&\leq & a\left(\eta, R(\omega_m(k)) \eta  \right) +
(aE^V(m,N)+b) \left(\eta, R(\omega_m(k))^2 \eta  \right) \nonumber \\
&\leq & C \frac{(\Lambda + 1)}{|k|^2}\hspace{2em} \mbox{for}\ |k|\leq\Lambda
\label{estimate2}
\end{eqnarray}
where $a$ and $b$ are independent of $m$. Since $\sup_{m <1}E^V(m,N) <
\infty$ the constant $C$ is also independent of $m$. Finally the
second factor in (\ref{estimate1}) is bounded by
$|k|\Vert |X| \widetilde{\Phi}_m\Vert$. 
The term containing the Pauli matrices in (\ref{anequation}) is estimated 
similarly. In conjuction with
(\ref{anequation}), (\ref{estimate0})(\ref{estimate1}) and (\ref{estimate2})
this shows that
\begin{equation}
\Vert \an(k)\Phi_m\Vert\leq
C\sqrt{\alpha}(\Lambda+1)^{1/2}\frac{\chi_{\Lambda}(k)}{|k|^{1/2}} 
\Vert |X| \widetilde{\Phi}_m\Vert
\end{equation}
This, together with Lemma \ref{exp},  proves the theorem.
\end{proof}

Next we differentiate (\ref{anequation}) with respect to $k$. 
There is a slight problem
with this calculation since the polarization vectors cannot be defined
in a smooth fashion globally. 
We make the following choice for the polarization vectors.
\begin{equation}\label{eq:pol1}
\varepsilon_{1}(k)= \frac{( k_2,-k_1,0 )}
{\sqrt{k_1^2+k_2^2}}
\end{equation}
and
\begin{equation}\label{eq:pol2}
\varepsilon_{2}(k) = \frac{k}{|k|} \wedge \varepsilon_{1}(k) \ .
\end{equation}

\begin{thm}[Photon derivative bound]\label{finitederiv}
Assume that there is binding, i.e., $\Sigma^V(m,N) - E^V(m,N) > 0$.
Assume that $\Phi_m$ is a normalized ground state for the many-body
Hamiltonian
$H^V_m $, $m \geq 0$. Then for $|k| < \Lambda$ and $(k_1,k_2) \not= (0,0)$
\begin{equation} \label{cruxbound}
\Vert \nabla_{k} \an(k) \Phi_m \Vert < \frac{Q \sqrt{\alpha}(1+|g|)}{|k|^{1/2}
\sqrt{k_1^2 + k_2^2}},
\end{equation}
where $Q$ is a finite constant independent of $\Phi_m$,  $g$,
$\alpha$, $\Lambda$, and depends  on $m$ only through
the binding energy $\Sigma^V(m,N)-E^V(m,N) >0$ .
\end{thm}

\begin{proof} We differentiate \ref{anequation} with respect to $k$
and obtain
\begin{eqnarray*}
& & \nabla_{k}(\an(k)) \widetilde{\Phi}_m = \nonumber \\
& & 2\sqrt{\alpha} R(\omega_m(k))^2 \frac{k}{|k|}
\varepsilon_\lambda(k)\cdot \sum_{j=1}^N
(p_j + \sqrt{\alpha}\, \widetilde{A}(x_j)) \frac{(1-e^{-ik\cdot x_j})}
{|k|^{1/2}}\widetilde{\Phi}_m  + \nonumber \\
& & 2\sqrt{\alpha} R(\omega_m(k))
\nabla_{k} ( \varepsilon_\lambda(k)) \cdot \sum_{j=1}^N
(p_j + \sqrt{\alpha}\, \widetilde{A}(x_j)) \frac{(1-e^{-ik\cdot x_j})}
{|k|^{1/2}}\widetilde{\Phi}_m  + \nonumber \\
& & 2\sqrt{\alpha} R(\omega_m(k))
 \varepsilon_\lambda(k) \cdot \sum_{j=1}^N
(p_j + \sqrt{\alpha}\,\widetilde{A}(x_j)) \nabla_{k} \left( \frac{(1-e^{-ik\cdot x_j})}
{|k|^{1/2}}\right)\widetilde{\Phi}_m  \\
& & +\frac{g}{2}\sqrt{\alpha}\, \nabla_{k} \left(i R(\omega_m(k)) 
\frac{k\wedge  \varepsilon_\lambda(k)}{\sqrt{|k|}} 
\cdot \sum_{j=1}^N \sigma_j e^{-ik\cdot x_j}\widetilde{\Phi}_m \right)
  \\ 
& & - \nabla_{k}\left(R(\omega_m(k))\omega_m(k)iw_{\lambda}(k,X)\widetilde{\Phi}_m \right) \ .
\end{eqnarray*}

The norms of the first and third terms can estimated precisely the same way
as in (\ref{estimate1})
and (\ref{estimate2}), and yields a bound of the form
\begin{equation}
\frac{C \sqrt{\alpha}}{|k|^{3/2}} 
\Vert \left( 1+ |X| \right) \widetilde{
\Phi}_m \Vert \ . 
\end{equation}
For the second term, a straightforward calculation shows that
\begin{equation}
|\nabla_k \varepsilon_{i}(k)| \leq \frac{{\rm const.}}{\sqrt{k_1^2 + k_2^2}}
\ \ {\rm for}\ \ i=1,2.
\end{equation}
The last term is dealt with in a similar fashion as the previous ones.
Using the steps in (\ref{estimate1}) and (\ref{estimate2}), 
this leads to the bound
\begin{equation}
\Vert \nabla_{k}(\an(k)) \widetilde{\Phi}_m \Vert 
\leq \frac{C \sqrt{\alpha}}{|k|^{1/2}\sqrt{k_1^2+k_2^2}} 
\Vert \left( 1+ |X| \right) \widetilde{ \Phi}_m \Vert \ . \label{derivbound}
\end{equation}
The fourth term can be estimated in the same fashion to yield a similar
result. 

Differentiating (\ref{relation}) leads to the same estimate with
$ \widetilde{\Phi}_m$ replaced by $\Phi_m$. This, together with
Lemma \ref{exp}, proves the theorem.
\end{proof}

As for the proof of Theorem \ref{finitephot}, our somewhat formal 
calculations above are rigorously justified
in Appendix \ref{sec:appendixb}.

\section{PROOF OF THEOREM \ref{t1} } \label{sec:proof}

The proof will be done in two steps.

\begin{proof}
{\bf Step 1.} The Hamiltonian $H^V_m$ has a normalized ground state
$\Phi_m$, by Theorem \ref{existence_with_mass}.
Pick a sequence $m_1>m_2>...$ tending to zero and denote the corresponding
eigenvectors  by $\Phi_j$. This sequence is a minimizing sequence for $H^V_0$
by Theorem \ref{minseq}. Since $\|\Phi_j\|$ is bounded there is a subsequence
(call it again $\Phi_j$) which has a weak limit $\Phi$. Since
$H^V_0-E^V(0,N)\geq 0$ and by the lower semi-continuity of non-negative
quadratic forms (in our case, $H^V_0-E^V(0,N)$)
\begin{equation*}
0 \leq
(\Phi,(H^V_0-E^V(0,N))\Phi)\leq\liminf_{j\rightarrow\infty}
(\Phi_j,(H^V_0-E^V(0,N))\Phi_j) =0\ .
\end{equation*}
Hence $\Phi$ will be a (normalized) ground state if we show that $\|\Phi\|=1$ (i.e.
$\Phi_j\rightarrow \Phi$ strongly).
It is important to note, however that if we write
$\Phi_j = \{\Phi_{j,0}, \Phi_{j,1}, ..., \Phi_{j,n}, ...\}$, where 
$\Phi_{j,n}$ is the $n$-photon component of $\Phi_{j}$ then it suffices
to prove the $L^2$ norm-convergence of each $\Phi_{j,n}$. The reason is
the uniform bound on the total average photon number; see the remark
after Theorem \ref{finitephot} which implies
$$ 
\sum_{n\geq
N}\Vert\Phi_{j,n}\Vert^2 \leq \mbox{const}\ N^{-1}  \   .
$$
Likewise, it suffices to prove the strong
$L^2$ convergence in the bounded domain in which $|X| < R$ for each 
finite $R$. The reason for this is the exponential decay given
in Lemma \ref{exp}, which is uniform by Lemma \ref{binding_with_mass}.  
Finally, by Theorem \ref{finitephot} $\Phi_{j,n}(X,k_1,\ldots,k_n)$
vanishes if $|k_i|>\Lambda$ for some $i$. So it suffices to show $L^2$
convergence for $\Phi_{j,n}$ restricted to
$$
\Omega = \{(X,k_1,\ldots,k_n):|X|<R;\ |k_i|<\Lambda,\ i=1,\ldots,n\}\subset \R^{3(N+n)}
$$
for each $R>0$.

{\bf Step 2.}  For each $p<2$ and $R>0$ we show that $\Phi_{j,n}$ restricted to
$\Omega$ is a bounded sequence in $W^{1,p}(\Omega)$.  The key to this bound is
(\ref{cruxbound}) and 
\begin{equation}\label{andefinition}
(\an(k)\Phi_j)_{n-1}(X,k_1,\ldots,k_{n-1}) =
\sqrt{n}\Phi_{j,n}(X,k,k_1,\ldots,k_{n-1}) 
\end{equation} where the arguments
$\lambda,\lambda_1,\ldots,\lambda_{n-1}$ and the spin indices have been
supressed.  By the symmetry of $\Phi_{j,n}$, (\ref{andefinition}), H\"older's
inequality and (\ref{cruxbound}) 
\begin{multline} \int_{B_R}dX\int_{|k_1|,
\dots, |k_n| <\Lambda} dk_1\ldots dk_n
\sum_{i=1}^n|\nabla_{k_i}\Phi_{j,n}(X,k_1,\ldots,k_{n})|^p\\ = 
n^{1-p/2}\int_{B_R}dX\int_{|k_1|, \dots, |k_n| <\Lambda} dk_1\ldots dk_n
|\nabla_{k_1}(\an(k_1)\Phi_{j})_{n-1}(X,k_2,\ldots,k_{n})|^p\\ \leq C
\int_{|k_1|<\Lambda} dk_1\left( \int_{B_R}dX\int_{|k_2|, \dots, |k_n| <\Lambda}
dk_2\ldots dk_n |\nabla_{k_1}(\an(k_1)\Phi_{j})_{n-1}(X,k_2,\ldots,k_{n})|^2
\right)^{p/2}\\ \leq C \int_{|k_1|<\Lambda}
dk_1\Vert\nabla_{k_1}\an(k_1)\Phi_{j}\Vert^p \leq \mbox{const} 
\end{multline} 
independent of $j$.  The constant $C$ depends on all the
parameters, but is finite because $|k_i|\leq\Lambda$ in the integration.
Similarly, by H\"older's inequality \begin{eqnarray*} \Vert
\chi(|X|<R)\nabla_X\Phi_{j,n}\Vert_p^p &\leq& C \Vert
\chi(|X|<R)\nabla_X\Phi_{j,n}\Vert_2^{p}\\ &\leq& C(\Phi_j,\sum_{i=1}^N p_i^2
\Phi_j)^{p/2} \end{eqnarray*} which is uniformly bounded 
by Lemma (\ref{lm:kinbound}).

Since the classical derivative of $a_{\lambda}^{\phantom *}(k) \Phi_m$ is not
defined in all of $\Omega$ one has to check that the weak derivative coincides
with the classical derivative a.e.. Because of our definitions (\ref{eq:pol1})
and (\ref{eq:pol2}), the classical derivative is
not defined along the $3$-axis.

One has to show that
$$
\int_{\Omega} \partial_i \psi \Phi_{j,n}  
= \lim_{\varepsilon \to 0}\int_{\Omega_\varepsilon} \partial_i \psi \Phi_{j,n}
= -\lim_{\varepsilon \to 0}\int_{\Omega_\varepsilon}\psi \partial_i \Phi_{j,n}
\ \ \ \  i=1, \dots, 3(N+n).
$$
for any test function $\psi \in C^{\infty}_c(\Omega)$. Here
$\Omega_{\varepsilon}$ is $\Omega$ with an $\varepsilon$ cylinder around the
$3$-axis removed in each $k$-ball. The first equality is trivial; it is the second equality that
has to be checked. This amounts to showing that the boundary term, coming
from the integration by parts vanishes in the limit as $\varepsilon$ tends
to zero. But this follows immediately from Theorem \ref{finitephot}.

This shows that $\Phi_{j,n}$ as a function of all its $3(N+n)$ variables, is in
the Sobolev space $W^{1,p} (\Omega)$ and that
$\sup_j\|\Phi_{j,n}\|_{W^{1,p}(\Omega)} <\infty$. Since $\Phi_{n,j}$ converges
weakly in $L^2(\Omega)$ it converges weakly in $L^p(\Omega)$ and since
the sequence is bounded in $W^{1,p} (\Omega)$, $\nabla \Phi_{n,j}$
converges weakly to $\nabla \Phi_{n}$.

The Rellich-Kondrachov theorem (see \cite{LL1} Theorem 8.9)
states that such a sequence converges {\it strongly} in $L^q(\Omega)$
if $1 \leq q < \left[\, 3p(N+n)/ 3(N+n)-p \,\right]$. The boundedness of $\Omega$ is crucial 
here. For our purposes we
need $q=2$, and hence we have to pick $p$ such that
\begin{equation}
2> p > \frac{2\cdot 3(N+n)}{2+3(N+n)}
\end{equation}
which is possible for each $N$ and $n$. We conclude that $\Phi_{j,n}\to \Phi_{n}$ strongly in 
$L^2(\Omega)$ as $j\to \infty$, for each $n$ and $R$. 
This proves the theorem. 
\end{proof}

{\it Remark:} Theorem \ref{finitederiv} essentially says that the derivative
is almost, but not quite in $L^2$. For high dimensions, $3(N+n)$, the
required $p$ is as close as we please to 2 if we require $q=2$, but $p=2$ is
not allowed. The way out of the difficulty was to prove a uniform bound on
the number operator and use this to say that that it suffices to prove strong
convergence for each $n$ separately. With $n$ then fixed, it is possible to
find a $p<2$ that yields $q=2$. It is, therefore, crucial to have the
derivative in every $L^p$ space with $p<2$.  The resolution of the problem of
the infrared singularity is thus seen to be a delicate matter. 

\appendix

\section{Appendix: LOCALIZATION ESTIMATES}  \label{sec:appendix}

In this appendix we collect a few facts which we use several times in this
paper. Generally we worry about localizations of Hamiltonians in
configuration space. While this is standard for Schr\"odinger operators it is
somewhat more complicated in the presence of the radiation field. This is
chiefly due to the problem of localization of photons.

We begin by stating a few well known facts about partitions of unity.  Let
$\beta$ denote one of the $2^N$ subsets of the set of integers
$1, 2, ..., N$.  Its complement is denoted by $\beta^{c}$.  As shown in
\cite{HuSi} there exists a family of smooth functions $j_{\beta}$ 
having the following four properties.

(i)
\begin{equation}
 \sum_{\beta} j_{\beta}^2 = 1 \ .
\end{equation}

(ii) For $\beta \not= \{1, \dots, N \}$ (including the empty set) the
$j_{\beta}$'s are homogeneous of degree $0$ and live outside the ball of
radius $R$
centered at the origin and
\begin{equation}
 \supp j_{\beta} \subset \{X : \min_{i \in \beta, j \in \beta^c}
(|x_i-x_j|, |x_j|) \geq c|X| \} \ ,
\end{equation}
where $C$ is some positive constant.

(iii) In the case where $\beta = \{1, \dots, N\}$,  $j_\beta$ is compactly
supported.

Corresponding to these electron localizations we define  photon
localizations.

For given $\beta \not= \{1, \dots, N \}$ consider the function
\begin{equation}
g_1(y;\beta,X) = \Pi_{j \in \beta^c} \left( 1- \chi(\frac{y-x_j}{P}) \right)
\end{equation}
where $\chi$ is a smoothed characteristic function of the unit ball.
Define $g_2(y;\beta,X) = 1 -g_1(y,\beta,X)$.
In the variable $y$, the function $g_1$ is supported away from the particles 
in $\beta^c$ while $g_2$ lives close to the particles in $\beta^c$.
Next define, for $i=1,2$,~~~~~~~ 
\begin{equation}
j_i(y;\beta,X) = \frac{g_i(y;\beta,X)}{\sqrt{g_1(y;\beta,X)^2+
g_2(y;\beta,X)^2}} \ .
\end{equation}
Certainly $j_1^2+j_2^2 = 1$ and a simple computation shows that
\begin{equation}
|\nabla j_i | \leq \frac{{\rm const.}}{P} \ .
\end{equation}

In the case where $\beta = \{1, \dots, N \}$ the construction 
of $j_1$ and $j_2$
is similar
to the above one except that the function $g_1$ depends on $y$, 
is equal to one in a neighborhood of the origin and is compactly supported.

With the help of $j_1$ and $j_2$ the photons can now be localized as  was
done in Section 4. Let $U_{\beta}(X) {\cal F} \to {\cal F} \otimes {\cal F}$ 
be the corresponding isometric transformation,
i.e., the one that is defined via the relation
\begin{equation}
U_{\beta}(X) a^{\#}(h) U^*_{\beta}(X) = a^{\#}(j_1 h) \otimes {\cal I} +
{\cal I} \otimes a^{\#}(j_2 h) \ .
\end{equation} 
The tensor product indicated is a tensor product between Fock spaces.

We denote by $H_{\beta}$ the Hamiltonian  of the form (\ref{ham}) with
photon mass, but only for the particles in the set $\beta$. More precisely
this operator acts on $L^2(\R^{3|\beta|})\otimes {\cal F}$.  By
$H^{\beta^c}$ we denote the Hamiltonian of the form (\ref{ham}) with photon
mass, but only for the particles in the set $\beta^c$ where the interaction
with the nuclei has been dropped. This operator acts on
$L^2(\R^{3|\beta^c|})\otimes {\cal F}$ In particular we keep the interaction
among those particles.  In the case where $\beta = \{1, \dots, N \}$ the
Hamiltonian $H^{\beta^c} = H_f(m)$.
 
 \begin{lm}[Localization of Hamiltonian]\label{locallemma}
For every $\beta$ 
\begin{equation}
j_{\beta} H j_{\beta} = U^*_{\beta}(X) j_{\beta}
\left [ H_{\beta} \otimes {\cal I} + {\cal I} \otimes H^{\beta^c} \right]j_{\beta} 
U_{\beta}(X) + o(1)\ .
\end{equation}
For $\beta\not= \{1,...,N\}$, $o(1) \to 0$ as first $R \to \infty$ and then $P \to \infty$. 
If $\beta = \{1, \dots, N\}$ then $o(1) \to 0$ as $P \to \infty$
for every fixed $R>0$.
\end{lm}

\begin{proof}

Our immediate aim is to compare the field energy $H_f$ with the
localized field energy $U^*_{\beta}(X) [H_f \otimes {\cal I} + {\cal I} \otimes H_f]
U_{\beta}(X)$.
For simplicity the various indices are supressed and $U_{\beta}(X)$
is replaced by $U_{\beta}$. The variable $X$ plays no role here.
Pick an orthonormal basis $\{ g_j \}_{j=1}^{\infty}$ of $L^2(\R^3)$
in $H^{1/2}(\R^3)$.
States of the form
\begin{equation}
\zeta = const. a^*_{\lambda_{i_1}}(g_{i_1}) \cdots a^*_{\lambda_{i_k}}(g_{i_k}) |0>
\label{state}
\end{equation}
where $k$ is finite, form an orthonormal basis in the Fock space.
The field energy acts on such states as
\begin{equation}
H_f \zeta =\sum_{j=1}^k a^*_{\lambda_{i_1}}(g_{i_1})  \cdots
a^*_{\lambda_{i_j}}( \omega g_{i_j}) \cdots  a^*_{\lambda_{i_k}}(g_{i_k})
|0>\ .
\end{equation}
Thus, we have that
\begin{equation}
H_f \zeta = U^*_{\beta}\sum_{j=1}^k c^*_{\lambda_{i_1}}(g_{i_1})  \cdots
c^*_{\lambda_{i_j}}( \omega g_{i_j}) \cdots
c^*_{\lambda_{i_k}}(g_{i_k})
 U_{\beta} |0>
\end{equation}
and
\begin{equation}
H_f = U^*_{\beta} \left[ H_f \otimes {\cal I} + {\cal I} \otimes H_f \right] U_{\beta} 
+ E_f
\end{equation}
where the error $E_f$ is given by
\begin{equation}
\begin{split}
E_f \zeta
=U^*_{\beta} \sum_{j=1}^k c^*_{\lambda_{i_1}}(g_{i_1})  \cdots &
(a^*_{\lambda_{i_j}}([j_1,\omega] g_{i_j})\otimes {\cal I} + {\cal I}\otimes
a^*_{\lambda_{i_j}}([j_2,\omega] g_{i_j}))\\ &\cdots
c^*_{\lambda_{i_k}}(g_{i_k})U_{\beta} |0> \ .
\end{split}
\end{equation}
Thus $E_f$ is given by the operator (the $\lambda$'s are omitted)
\begin{align}
E_f = & U^*_{\beta} \sum_{k} \left[ a^*([j_1,\omega]g_k)\otimes {\cal I}
+ {\cal I} \otimes a^*([j_2,\omega]g_k) \right] 
\left[a^{\phantom {*}} (j_1 g_k)\otimes {\cal I}   + 
{\cal I} \otimes a^{\phantom{*}}(j_2g_k)\right]
U_{\beta} \ .
\end{align}
The expression for the operator $E_f$ does not look hermitian but it is,
remembering that $U^*_{\beta}$ is a partial isometry.
Standard estimates lead to
\begin{equation}\label{kparticle}
|\left( \Psi, E_f \Psi \right)|
\leq   \left( \Vert [\omega,j_1]\Vert +  \Vert [\omega,j_2] \Vert \right) 
\left(\Psi,\left[{\cal N}+1\right] \Psi \right)\ .
\end{equation} 
where ${\cal N}$ is the number operator.

Here $ \Vert[j_1, \omega] \Vert$ denotes the operator norm
associated with the kernel $[j_1, \omega]$.
This norm can be estimated using the formula
\begin{equation}
[j_1, \omega] = [j_1,\omega^2]\frac{1}{\omega} + \omega^2 [j_1, \frac{1}
{\omega}] \ .
\end{equation}
Recalling the definition of $j_1$, the operator norm of the
first term is easily seen to be bounded by a $const./P$.
Likewise, the second term, using the formula
\begin{equation}
\frac{1}{\sqrt{p^2+m^2}}= \frac{1}{\pi} \int_0^{\infty} \frac{1}
{t+p^2+m^2}\frac{dt}{\sqrt t}\ ,
\end{equation}
can be estimated by $const./P$. The term $\Vert [j_2, \omega] \Vert$
is estimated in a similar fashion.
The estimate (\ref{kparticle}) immediately shows that for a general
state $\Phi$ we have that
\begin{equation}
|(\Phi,\left[ H_f - U^*_\beta \left[ H_f \otimes {\cal I} + {\cal I} \otimes H_f \right]
 U_\beta \right] \Phi)| \leq \frac{const.}{P} (\Phi, {\cal N} \Phi)\ .
\end{equation}
Since the photons have a mass we
can estimate the number operator in terms of the field energy.
The field energy is relatively bounded with respect to the Hamiltonian,
i.e., $H_f \leq aH^V_m + b$ for some positive constants $a$ and $b$,
and thus we obtain
\begin{equation}
|(\Phi,\left[ H_f - U^*_\beta \left[ H_f \otimes {\cal I} + {\cal I} \otimes H_f \right]
 U_\beta \right] \Phi)| \leq \frac{ const.}{Pm}(\Phi, [aH^V_m+b] \Phi) \ .
\end{equation}
Note that this estimate had nothing to do with the electron, in
particular the $x$--space cutoff is not present in the calculation.

Next we have to compare 
$\sum_{j=1}^N (p_j+\sqrt{\alpha}\, A(x_j))^2$ with
$$
U^*_{\beta}(X) j_{\beta} 
\left[ \sum_{i \in \beta} (p_i+\sqrt{\alpha}\, A(x_i))^2\otimes {\cal I} + {\cal I} \otimes
\sum_{j \in \beta^c} (p_j+\sqrt{\alpha}\, A(x_j))^2 \right] j_{\beta} U_{\beta}(X)\ .
$$
This time the $X$--space cutoff is important. 
We would like to estimate the difference
\begin{eqnarray}
& & j_{\beta} \sum_{i \in \beta} \left[(p_i+\sqrt{\alpha}\, 
A(x_i))^2 - U^*_{\beta}(X)  
(p_i+\sqrt{\alpha}\, A(x_i))^2 \otimes {\cal I} U_{\beta}(X)\right] j_{\beta} 
\nonumber \\ 
&+& j_{\beta} \sum_{i \in \beta^c} \left[(p_i+\sqrt{\alpha}\, A(x_i))^2 - 
U^*_{\beta}(X)  
{\cal I} \otimes (p_i+\sqrt{\alpha}\, A(x_i))^2  U_{\beta}(X) \right] j_{\beta}\ .
 \label{decomp}
\end{eqnarray}
It suffices to treat the first term, the other is similar.
It can be easily expressed as
\begin{equation}
 j_{\beta} \left[\sum_{i \in \beta}
 (p_i+\sqrt{\alpha}\, A(x_i))Q_i+Q_i(p_i+\sqrt{\alpha}\, A(x_i)) -
Q_i^2\right] j_{\beta}
\end{equation}
where
\begin{equation}
Q_i=p_i+\sqrt{\alpha}\, A(x_i)- U^*_{\beta}(X)(p_i+ \sqrt{\alpha}\, 
A(x_i)) \otimes {\cal I} U_{\beta}(X)\ .
\end{equation}
Using the form boundedness of the kinetic energy
with respect to the full Hamiltonian, we have
\begin{equation}
(\Psi,\sum_{j=1}^N (p_j+\sqrt{\alpha}\, A(x_j))^2 \Psi)
 \leq a(\Psi, H^V_m \Psi) +b(\Psi, \Psi)
\end{equation}
for positive constants $a$ and $b$.. Thus, using Schwarz' inequality
it suffices to show that
\begin{equation}
\Vert  Q_i j_{\beta}\Psi \Vert = o(1)\ \ {\rm for}\ \ i \in \beta \ ,
\end{equation}
as $R$ (the localization radius for the electrons) tends to infinity.
Denote by
\begin{equation}
h^{\lambda}_{i,x}(y) = (2 \pi)^{-3/2} \int_{|k| < \Lambda}
\frac{1}{\omega(k)}   \varepsilon_i^{\lambda}(k) e^{i k\cdot (y-x)} d^3 k \ .
\end{equation}
Explicitly, $Q_i$ is given by
\begin{eqnarray}
& & p_i - U^*_{\beta}(X) p_i \otimes {\cal I} U_{\beta}(X) \nonumber \\
& + & U^*_{\beta}(X) \left[ \sum_{\lambda} \an([j_1-1]h^{\lambda}_{x})
\otimes {\cal I} + {\cal I}\otimes \an(j_2h^{\lambda}_{x})\right] U_{\beta}(X) \nonumber \\
& + & U^*_{\beta}(X) \left[ \sum_{\lambda} \cre([j_1-1]h^{\lambda}_{x})
\otimes {\cal I} + {\cal I}\otimes \cre(j_2 h^{\lambda}_{x}) \right] U_{\beta}(X) \ , 
\label{difference}
\end{eqnarray}
and it suffices to estimate each of these terms separately. Each of the 
last two terms can be brought into the form
\begin{equation}
\Vert U^*_{\beta}(X) a^{\#}(f) \otimes {\cal I} U_{\beta}(X) 
j_{\beta}\Psi \Vert \label{small}
\end{equation}
where $f$ is one of the functions
\begin{equation}
[j_1(y,\beta,X)-1]h^{\lambda}_{1,x_j}(y)\ \ {\rm or}\ \  
j_2(y,\beta,X)h^{\lambda}_{1,x_j}(y)\qquad\qquad  j \in \beta\ .
\end{equation}
The terms (\ref{small}) are estimated by
\begin{equation}
\sup_{X}\{j_{\beta}(X) \Vert [j_1 -1] h_{i,x_j}^{\lambda}\Vert_2 \}
\sqrt{(\Psi, ({\cal N}+1) \Psi)} \label{overlap1}
\end{equation}
respectively
\begin{equation}
\sup_{X}\{j_{\beta}(X) \Vert j_2 h_{1,x_j}^{\lambda}\Vert_2 \}
\sqrt{(\Psi, ({\cal N}+1) \Psi)} \ . \label{overlap2}
\end{equation}
In both formulas the index $j$ is in $\beta$.
The function $j_{\beta}$ lives in the region where $|x_i - x_j| \geq cR$
for $i \in \beta$ and $j \in \beta^c$. The function $j_1 -1$
(and likewise $j_2$)
is not zero only if $|y - x_j| \leq P$ for some $j \in \beta^c$. 
Thus, $j_{\beta}(X)(j_1-1)(y)$ and $j_{\beta}(X)j_2(y)$ are nonzero only
if $|y-x_i| \geq cR - P$.
As $cR-P$ gets large only the tail of the function $h^{\lambda}$
contributes to the integral which can be made as small as we please.
The number operator is bounded by the field energy times
$1/m$ which in
turn is bounded by the full energy. 

To estimate the first term in (\ref{difference}) we calculate
\begin{equation}
\begin{split}
& p_i - U^*_{\beta}(X) p_i \otimes {\cal I} U_{\beta}(X) = \\
& U^*_{\beta}(X) \sum_{k} \left[ a^*([p_i,j_1]g_k)\otimes {\cal I}
+ {\cal I} \otimes a^*([p_i,j_2]g_k) \right] \times \\
& \left[a^{\phantom {*}}
(j_1 g_k)\otimes {\cal I}   + {\cal I} \otimes a^{\phantom{*}}(j_2g_k)\right]
U_{\beta}(X) \ .
\end{split}
\end{equation} 
Note that the tensor product in the first line is different from the second.
In the first the identity acts on $L^2(\R^{3|\beta^c|})\otimes{\cal F}
\otimes{\cal F}$ while in the second $\otimes$ indicates the tensor product
of the Fock spaces only. The functions $g_k$ indicates a basis of $L^2(\R^3)$.
The operators $U^*_{\beta}(X)$ and $U_{\beta}(X)$ have unit norm. Thus
\begin{equation}
\begin{split}
& \Vert U^*_{\beta}(X) \sum_{k} \left[ a^*([p_i,j_1]g_k)\otimes {\cal I}
+ {\cal I} \otimes a^*([p_i,j_2]g_k) \right] \times  \\  
&\left[a^{\phantom {*}} (j_1 g_k)\otimes {\cal I}   + 
{\cal I} \otimes a^{\phantom{*}}(j_2g_k)\right]
U_{\beta}(X) \Psi \Vert  \\
& \leq   \left( \Vert [p_i,j_1]\Vert +  \Vert [p_i,j_2] \Vert \right) 
\Vert \sqrt{{\cal N}+1} \Psi\Vert\ ,
\end{split}
\end{equation}
where  $\Vert \cdot \Vert$ indicates that
the operator norm has been taken. The norms of the
commutators are of the order $1/P$ and
hence vanish as $P \to \infty$. Since the photons have a mass we
can estimate the number operator in terms of the field energy.

Similar consideration apply to  the $\beta^c$ term in (\ref{decomp}).
The only difference is that instead of  (\ref{overlap1}) and (\ref{overlap2})
we have 
\begin{equation}
\sup_{X}\{j_{\beta}(X) \Vert j_1 h_{x_j}^{\lambda}\Vert_2 \}
\sqrt{(\Psi, ({\cal N}+1) \Psi)} 
\end{equation}
respectively
\begin{equation}
\sup_{X}\{j_{\beta}(X) \Vert [j_2-1] h_{x_j}^{\lambda}\Vert_2 \}
\sqrt{(\Psi, ({\cal N}+1) \Psi)} \ , 
\end{equation}
with $j \in \beta^c$. Again this terms tend to zero as $P \to \infty$.
The proof for the case where $\beta = \{1, \dots, N \}$ is similar
but simpler since the operator $U_{\beta}$ does not depend on $X$.

Finally, we have to compare the $\sigma \cdot B$ term with its localized
counterparts. The estimates  are similar to, but much easier than the
estimates for $(p+\sqrt\alpha \, A(x))^2$ and are omitted for the convenience
of the reader and authors who, by now, are exhausted.  

\end{proof}

A simple consequence of Lemma \ref{locallemma} is the following.
\begin{cl}\label{threshold}
Let $\phi$ be a smooth function on $\R^{3N}$ such that $j_{\beta} \phi 
\equiv 0$
for $\beta = \{1, \dots , N \}$. Thus,  $\phi$ depends on $R$.
Then, as operators,
\begin{equation}
\phi H \phi \geq \left( \Sigma^V(m,N) + o(1) \right) \phi^2 \ .
\end{equation}
Here, $\Sigma^V(m,N) = \min_{1 \leq N' <N}(E^V(N') + E^0(N-N'))$
and $o(1)$ vanishes as $R \to \infty$.
\end{cl}

\begin{proof} By the IMS localization formula we have that
\begin{equation}
\phi H \phi = \sum_{\beta} \phi j_{\beta}H j_{\beta}\phi  - 
\phi^2 \sum_{\beta} |\nabla j_{\beta}|^2 \ ,
\end{equation}
where the second term goes to 0 as $R \to \infty$.
With our assumption on $\phi$ only the sets $\beta$ with $\beta^c \not= 
\emptyset$ contribute. {F}rom Lemma \ref{locallemma} we get that
\begin{equation}
\phi H \phi =\sum_{\beta} U^*_{\beta}(X)\phi j_{\beta}
\left [ H_{\beta} \otimes {\cal I} + {\cal I} \otimes H^{\beta^c} \right]j_{\beta}\phi 
U_{\beta}(X) + o(1)
\end{equation}
as first $R \to \infty $ then $P \to \infty$. 
Certainly $H_{\beta} \geq E^V(m,|\beta|)$ and
$H^{\beta^c} \geq E^0(m,|\beta^c|)$ from which the statement immediately 
follows.
\end{proof}

\begin{lm}\label{compactlemma} Let $\Psi_n$  be a normalized sequence in
${\cal H}$ whose energy is uniformly bounded and
such that for any $\Phi \in {\cal H}$ with finite energy,
\begin{equation}
(\Psi_n, \Phi) \to 0\ ,\ {\rm and}\ \ (\Psi_n, H \Phi) \to 0 \ .
\end{equation}
Then
\begin{equation}
(\phi_R \Psi_n, U^* {\cal I} \otimes P_2 U \phi_R \Psi_n) \to 0 \ .
\end{equation}
Here $U$ is the Fock space localization $U_{\beta}$ that corresponds to
$\beta = \{1,...,N\}$. 
\end{lm}

\begin{proof}

Since the energy of $\Psi_n$ is uniformly bounded we also know
that
\begin{equation}
(\Psi_n, H^0(m) \Psi_n) \leq C \ .\label{enbound}
\end{equation}
is uniformly bounded

Let us describe the operator ${\cal I} \otimes P_2 U$ in more detail.
Recall that
\begin{equation}
U a^*(h_{i_1}) \cdots a^*(h_{i_k})|0> =
c^*(h_{i_1}) \cdots c^*(h_{i_k}) U |0>
\end{equation}
where $|0>$ denotes the vacuum vector in Fock space and
$U|0>=|0>\otimes|0>$. Hence, using the definition of $c^*(h)$, we find
that
\begin{equation}
{\cal I} \otimes P_2
Ua^*(h_{i_1}) \cdots a^*(h_{i_k}) |0> 
= a^*(h_{i_1}j_1) \cdots a^*(h_{i_k}j_1)|0>\otimes|0> \ .
\end{equation}
The projection $P_2$ annihilates the photons in the second factor.
In other words, the operator ${\cal I} \otimes P_2 U$ when acting on a state
\begin{equation}
\Psi = \{ \Psi^0, \Psi^1(y_1) , \Psi^2 (y_1,y_2), \cdots \}
\end{equation}
produces the localized state $\Gamma(j_1) \Psi\otimes |0>$ where
\begin{equation}
\Gamma(j_1) \Psi = \{ \Psi^0, j_1(y_1) \Psi^1(y_1),j_1(y_1)j_1(y_2)
\Psi^2(y_1,y_2)
, \cdots \} \ .
\end{equation}
It follows that
\begin{align*}
(\phi_R \Psi_n, U^* {\cal I} \otimes P_2 U \phi_R \Psi_n) &= \Vert {\cal I}
\otimes P_2 U \phi_R \Psi_n\Vert^2\\ &= \Vert\Gamma(j_1)\phi_R\Psi_n\Vert^2
\end{align*}

Next, we show that (\ref{enbound}) implies that
\begin{equation}
 \Gamma(j_1)\phi_R\Psi_n \to 0 \ .\label{zero}
\end{equation}

To achieve that we note first that on account of the positive mass
we have that $(\Psi_n, {\cal N} \Psi_n)$ is uniformly bounded.
Since $\Psi_n$ is of the form
$$
\left\{\Psi_n^0, \Psi_n^1(X, y_1), \Psi_n^2(X, y_1,y_2), \cdots \right\}
$$
we know that $\sum_{k\geq M}\left( \Psi_n^k , \Psi_n^k \right) \leq \mbox{const}/M$.
It is therefore sufficient to prove (\ref{zero}) for each function
$$
\Psi_n^M(X, y_1, \cdots, y_M) \ .
$$

{}From the lemma below we learn that
\begin{equation}
\sum_{j=1}(\Psi_n,p_j^2 \Psi_n)
\end{equation}
is uniformly bounded.
Thus, we can write (\ref{zero}) as
\begin{equation}
\Gamma(j_1)\phi_R(1 +\sum_{j=1} p_j^2 +
H_f )^{-1/2}(1 +\sum_{j=1} p_j^2  + H_f )^{1/2} \Psi_n \ .
\end{equation}
which vanishes as $n \to \infty$ since
$\Vert(1 +\sum_{j=1} p_j^2  + H_f )^{1/2} \Psi_n \Vert$
is uniformly bounded and since
$$
\Gamma (j_1) \phi_R(1 +\sum_{j=1} p_j^2 + H_f )^{-1/2}
$$
is compact on every finite particle subspace.  Compactness follows from
the fact that for continuous functions $f$ and $g$ vanishing at
infinity the operator $f(i\nabla)g(x)$ is compact.
\end{proof}

\begin{lm}[Bound on $A(x)^2$]\label{lm:Abound}
For each $x\in \R^3$ and ultraviolet cutoff
$\Lambda$
write $A(x) = D(x) +D^*(x)$ where $D$ contains the annihilation operators
in $A(x)$ and $D^*$ the creation operators. Similarly, write
$B(x) =E(x) +E^*(x)$. As operator bounds  
\begin{eqnarray}\label{eq:Abound}
H_f  &\geq &\frac{1}{8 \pi \Lambda}D^*(x)D(x)  \nonumber \\
H_f +\frac{\Lambda}{2} &\geq &\frac{1}{8 \pi \Lambda}D(x)D^*(x)  \nonumber \\
H_f +\frac{\Lambda}{8} &\geq & \frac{1}{32 \pi \Lambda}A(x)^2 \nonumber \\
H_f  &\geq  &\frac{3}{8 \pi \Lambda^3}E^*(x)E(x)\nonumber \\
H_f +\frac{3\Lambda}{4} &\geq  &\frac{3}{8 \pi \Lambda^3}E(x)E^*(x)\nonumber \\
H_f +\frac{3\Lambda}{16} &\geq  &\frac{3}{32 \pi \Lambda^3}B(x)^2\ . 
\end{eqnarray}
\end{lm}

\begin{proof} 
We write $A(x) = D(x) +D^*(x)$ with
$D(x) = \sum_\lambda \int_{|k|<\Lambda} |k|^{-1/2}\varepsilon_\lambda(k)
\exp[ik\cdot x]\an(k)d^3k$. There are thus four terms in
$A(x )^2 $ . Using the Schwarz inequality,
 the $(DD )$ term can be bounded above by $(D^* D  )/2 + (DD^* )/2$.
On the other hand, $(DD^* ) =(D^*D )+ \Gamma$, where
$\Gamma$ is the commutator $\int 2/|k| = 4\pi \Lambda^2 $;
the factor 2 comes from the two
polarizations $\lambda =1,2$. Altogether, we obtain
$$
 A(x )^2\leq
4 D^*(x)D(x)  + 4\pi \Lambda^2 \  .
$$
Finally, we use the Schwarz inequality again to obtain 
$$
\sum_\lambda \int \overline{h_\lambda(k)}\cre(k) d^3k\ \sum_\lambda 
\int h_\lambda(k)
\an(k) d^3k \leq \sum_\lambda \int |h_\lambda (k)|^2 /|k| d^3k \sum_\lambda
\int |k|\cre(k) \an(k) d^3k.
$$
In our case, $h_\lambda(k) = \varepsilon_\lambda(k)\exp[ik\cdot x]/\sqrt|k|$,
 so  $\sum_\lambda \int |h_\lambda (k)|^2 /|k| d^3k = 8\pi \Lambda $.

For  $B ={\rm curl}A$, 
replace $\Gamma$ by $2\pi \Lambda^4$ and 
replace $|h_\lambda (k)| $ by 
$ \sqrt{|k|}$.
\end{proof}

As a corollary of Lemma \ref{lm:Abound} we have the following.

\begin{lm} [Bound on $(p+A(x))^2$]\label{lm:kinbound} 
For any $\varepsilon >0$ there are   constants 
$\delta(\varepsilon) >0$ and $ C(\varepsilon) <\infty$ such that
\begin{equation}\label{eq:pbound}
\sum_{j=1}^N \left\{(p_j +\sqrt{\alpha}\, A(x_j))^2 + \frac{g}{2}
\sqrt{\alpha}\, \sigma_j\cdot
B(x_j) \right\} +\varepsilon H_f \geq  \delta(\varepsilon) 
\sum_{j=1}^N p_j^2  -  C(\varepsilon) \  .
\end{equation}
The constants $\delta(\varepsilon),C(\varepsilon) $ depend on 
$\alpha, \  g ,\   \Lambda, \   N$.
\end{lm}

\begin{proof}
In addition to Lemma \ref{lm:Abound}, use the facts that for any
$0<\mu,\nu<1$,
$(p_j +\sqrt{\alpha}\, A(x_j))^2 \geq (1-\mu)p^2 +(1- 1/\mu)\alpha A(x_j)^2$
and $2 \sigma_j\cdot B(x_j) \geq - \nu B(x_j)^2   - 1/\nu$.
\end{proof}

\section{Appendix: VERIFICATION OF INFRARED BOUNDS}
\label{sec:appendixb}

The proofs of the infrared bounds in Section~\ref{sec:infrared} are somewhat
formal. In particular, we carried out the calculations tacitly
assuming that $a_{\lambda}(k) \widetilde{\Phi}_m$ (which is itself only
defined for almost every $k$) is in the domain of the Operator
$\widetilde{H}_m$. One can actually prove this when
$\widetilde{H}_m$ is self-adjointly realized in terms of the
Friedrichs' extension and thereby make all the formal
computations in Section~\ref{sec:infrared} rigorous. Instead of doing so, we give
here alternative proofs of the Theorems in
Section~\ref{sec:infrared} which avoid any reference to a
domain of $\widetilde{H}_m$. All the arguments can be carried out
on the level of quadratic forms.

We recall that $\widetilde{H}_m$ is the Hamiltonian $H^V_m$ after an
``operator-valued gauge transformation''. Our remarks here about quadratic
forms in relation to $\widetilde{H}_m$ could just as well be applied
to $H^V_m$ itself.

In order to keep the notation simple, we give
the proof of the infrared bounds for the case of a single charged particle ($N=1$)
with no magnetic moment, i.e., $g=0$. There is no difficulty in deriving these
bounds for the general case.

Denote by ${\cal S}$ the set of all finite linear combinations of
vectors that are products of $C^{\infty}_c(\mathbb{R}^3)$-functions and states in
${\cal F}$ that have only a finite number of photons. It is well known
that this set is dense in ${\cal H}$, and that the quadratic form $(\Psi,\Psi)_+
:=(\Psi,(\widetilde{H}_m-E^V(m,1)+1) \Psi)$ is defined for all
$\Psi$ in ${\cal S}$ and is bounded below by $\| \Psi \|^2$.
Hence this quadratic form is closable and the closure of ${\cal
S}$ in this inner product is a Hilbert space $Q(\widetilde{H}_m)$
with inner product $(\cdot,\cdot)_+$ and norm $\Vert \Psi \Vert_+ =\sqrt{ (\Psi,\Psi)_+}$.

An eigenfunction $\widetilde{\Phi}_m$ of $\widetilde{H}_m$ in the weak sense
is a vector in $Q(\widetilde{H}_m)$ such that
\begin{equation}
(\Psi,\widetilde{\Phi}_m)_+ = e(\Psi, \widetilde{\Phi}_m)
\end{equation}
for some real number $e$ and for all $\Psi \in
Q(\widetilde{H}_m)$. It is in this sense that we proved in Section~
\ref{sec:mass} 
that a ground state exists for the model with massive photons. (This 
implies that $\widetilde{\Phi}_m$ is in an eigenstate of the Friedrichs'
extension of $\widetilde{H}_m$). 

Define the smeared operators
\begin{equation}
a(f)= \sum_{\lambda} \int a_{\lambda}(k) \overline{f(k, \lambda)} {\rm d} k \ ,
\end{equation}
where $f(k, \lambda)$ is any function in $L^2(\mathbb{R}^3;\mathbb{C}^2)$.  
It is not difficult to show that $a(f)\widetilde{\Phi}_m$
is in the form domain of $\widetilde{H}_m$. To this end define
$a_R(f) = R[{\cal N}+R]^{-1}a(f)$. Here $R$ is some large real number
(which we eventually take towards infinity) and ${\cal N}$ is the number
operator.

It is straightforward to see that $a_R(f)$ and $a^*_R(f)$
are bounded operators on $Q(\widetilde{H}_m)$ for every $R>0$,
i.e.,
\begin{equation}
\|a_R(f)\Psi\|_+ \leq C(R) \|\Psi\|_+ \ ,
\end{equation}
and similarly for $a^*_R(f)$.

Generally, the constant $C(R)$ tends to $\infty$ as $R$
tends to $\infty$. For an eigenfunction of $\widetilde{H}_m$,
however, this is not the case. Simple but tedious commutator
estimates reveal that for any eigenfunction $\widetilde{\Phi}_m$ there exists a
constant $C$ independent of $R$ such that
\begin{equation}
(a_R(f)\widetilde{\Phi}_m, a_R(f)\widetilde{\Phi}_m)_+ 
\leq C (a_R(f)\widetilde{\Phi}_m,a_R(f)\widetilde{\Phi}_m)
\ .
\end{equation}
The point is that $(a_R(f)\widetilde{\Phi}_m, a_R(f)\widetilde{\Phi}_m)_+ =
(a^*_R(f)a_R(f)\widetilde{\Phi}_m, \widetilde{\Phi}_m)_+ $ plus terms that are uniformly
bounded in $R$. By the previous statement we know that
$a^*_R(f)a_R(f)\widetilde{\Phi}_m$ is in $Q(\widetilde{H}_m)$ and hence
\begin{equation}
(a^*_R(f)a_R(f)\widetilde{\Phi}_m, \widetilde{\Phi}_m)_+ = 
e(a^*_R(f)a_R(f)\widetilde{\Phi}_m,
\widetilde{\Phi}_m) = e(a_R(f)\widetilde{\Phi}_m, a_R(f)\widetilde{\Phi}_m) \ . 
\label{afbound}
\end{equation}
The last expression, however, is  bounded uniformly in $R$, since
the condition $\widetilde{\Phi}_m \in Q(\widetilde{H}_m)$ implies that the
expectation value of the field energy in $\widetilde{\Phi}_m$ is finite which
in turn bounds the last expression in (\ref{afbound}). Here we
use the fact that the photons have a mass.

{}From this it follows easily that for a subsequence  of $R$'s tending to
infinity, $a_R(f)\widetilde{\Phi}_m$
has a weak limit in $Q(\widetilde{H}_m)$. Since
\(a_R(f)\widetilde{\Phi}_m\to a(f)\widetilde{\Phi}_m\) strongly this shows that
$a(f)\widetilde{\Phi}_m \in Q(\widetilde{H}_m)$. 

\bigskip
\begin{proof}[Proof of Theorem \ref{finitephot}]
We shall use the abreviation
\begin{equation}
\sum_{\lambda} \int \dots {\rm d}k =  \Sigma \kern-10pt \int \dots {\rm d} k\ .
\end{equation}
For our special choice of gauge
\begin{equation}
\widetilde{A}^i(x)=  a(G^i)+ a^*(G^i),  \ i=1,2,3\ ,
\end{equation}
where we set
\begin{equation}
G^i_{\lambda}(k,x) =  \varepsilon_{\lambda}^i(k)|k|^{-1/2}(e^{ik \cdot
x}-1)\chi_{\Lambda}(k) \ .
\end{equation}
Next, pick any $\Psi$ in ${\cal S}$ and calculate 
(recalling the definition of $w$ in Section
\ref{sec:infrared} equation (\ref{beeop}))
\begin{equation}
(\Psi, (\widetilde{H}_m-E^V(m,1))a(f) \widetilde{\Phi}_m) =-2(\Psi,
(f,G)(p+\widetilde{A}) \widetilde{\Phi}_m) - 
(\Psi, a(\omega f) \widetilde{\Phi}_m)+
i(\Psi, (f,\omega w)\widetilde{\Phi}_m)
\end{equation}
with $\omega(k) = \sqrt{k^2 + m^2}$. 
This extends, using an approximation argument, to all $\Psi \in
Q(\widetilde{H}_m)$ and, in particular, to $a(f) \widetilde{\Phi}_m$. 
Here we note
that, on account of Lemma \ref{lm:Abound} and the assumption on
the potential, $\Psi \in  Q(\widetilde{H}_m)$ implies that
$(p+\widetilde{A})\Psi \in \cal{H}$. Hence
\begin{eqnarray}
& 0 &\leq (a(f)\widetilde{\Phi}_m, (\widetilde{H}_m-E^V(m,1)) 
a(f) \widetilde{\Phi}_m) \\ \nonumber
& = & -2(a(f) \widetilde{\Phi}_m, (f,G)(p+\widetilde{A}) 
\widetilde{\Phi}_m) - (a(f) \widetilde{\Phi}_m,
a(\omega f) \widetilde{\Phi}_m) + 
i(a(f)\widetilde{\Phi}_m, (f,\omega w)\widetilde{\Phi}_m) \ ,
\end{eqnarray}
which yields the inequality
\begin{equation}
(a(f) \widetilde{\Phi}_m, a(\omega f) 
\widetilde{\Phi}_m) \leq 
-2(a(f)\widetilde{\Phi}_m,
(f,G)(p+\widetilde{A}) \widetilde{\Phi}_m) \label{keyineq}
+ i(a(f)\widetilde{\Phi}_m, (f,\omega w)\widetilde{\Phi}_m)
\end{equation}
for all $f$ in $L^2(\mathbb{R}^3;\mathbb{C}^2)$.  Pick $f$ of the form $\omega (k)^{-1/2}
q(k,\lambda) g_i(k,\lambda)$ where $g_i$ is an orthonormal basis of 
$L^2(\mathbb{R}^3;\mathbb{C}^2)$ and  $q(k,{\lambda})$ a
bounded function. Summing over this basis, we get on the left side of 
(\ref{keyineq}) 
\begin{equation}
\sum_i (a(\omega^{-1/2}q g_i) \widetilde{\Phi}_m, a(\omega^{1/2}q
g_i) \widetilde{\Phi}_m) = \Sigma \kern-10pt \int|q(k,\lambda)|^2 
\|a_{\lambda}(k)\widetilde{\Phi}_m\|^2  {\rm d}k \ ,
\end{equation}
and on the right side
\begin{equation}
  -2 (a(\omega^{-1}|q|^2G) \widetilde{\Phi}_m,(p+\widetilde{A}) 
  \widetilde{\Phi}_m) + 
  i(a(|q|^2 w)\widetilde{\Phi}_m, \widetilde{\Phi}_m)\ .
\end{equation}
Hence
\begin{equation}
\Sigma \kern-10pt \int |q(k,\lambda)|^2 \|a_{\lambda}(k)\widetilde{\Phi}_m\|^2 {\rm d}k \leq -2
(a(\omega^{-1}|q|^2G) \widetilde{\Phi}_m,(p+\widetilde{A}) 
\widetilde{\Phi}_m) +
i(a(|q|^2 w)\widetilde{\Phi}_m, \widetilde{\Phi}_m)\ .
\end{equation}
The right side can be written as
\begin{equation}
-2\Sigma \kern-10pt \int \frac{|q(k,\lambda)|^2}{\omega (k)}
(a_{\lambda}(k)\widetilde{\Phi}_m,G_{\lambda}(k)(p+\widetilde{A}) 
\widetilde{\Phi}_m) {\rm d}k +i\Sigma \kern-10pt \int |q(k,\lambda)|^2(a_{\lambda}(k) 
\widetilde{\Phi}_m, 
w_{\lambda}\widetilde{\Phi}_m) {\rm d} k \ ,
\end{equation}
and, applying Schwarz's inequality, this is bounded above by
\begin{eqnarray}
& 2 &  \left[ \Sigma \kern-10pt \int  |q(k,\lambda)|^2 \Vert a_{\lambda}(k) 
\widetilde{\Phi}_m\Vert^2{\rm d}k 
\right]^{1/2} \times \\ \nonumber 
& & \left[ \left[\Sigma \kern-10pt \int_{|k| \leq \Lambda} \omega(k)^{-2}
|q(k,\lambda)|^2\Vert G_{\lambda}(p+\widetilde{A}) \widetilde{\Phi}_m \Vert^2 {\rm d}k 
\right]^{1/2}
+ \left[\Sigma \kern-10pt \int_{|k| \leq \Lambda} |q(k,\lambda)|^2 \Vert 
w_{\lambda}\widetilde{\Phi}_m \Vert^2 {\rm d}k  \right]^{1/2} \right]
\ .
\end{eqnarray}
Hence we obtain the bound
\begin{eqnarray}
& &\Sigma \kern-10pt \int |q(k,\lambda)|^2 \|a(k)\widetilde{\Phi}_m\|^2 {\rm d}k \leq \\ \nonumber
& & 8 \sum_{\lambda} \left[ \Sigma \kern-10pt\int_{|k|
\leq \Lambda} \omega(k)^{-2} |q(k,\lambda)|^2\Vert
G_{\lambda}(k)\cdot(p+\widetilde{A}) \widetilde{\Phi}_m \Vert^2 {\rm d}k
+  \Sigma \kern-10pt \int_{|k| \leq \Lambda} |q(k,\lambda)|^2 \Vert 
w_{\lambda}\widetilde{\Phi}_m \Vert^2 {\rm d}k \right] \ .
\end{eqnarray}
Since, $\mathrm{div}_x \,G_{\lambda}=0$ we have that $G_{\lambda}
\cdot(p+\widetilde{A}) \widetilde{\Phi}_m = (p+\widetilde{A})\cdot G_{\lambda}
\widetilde{\Phi}_m$. Moreover, $(p+\widetilde{A})^2$ is relatively form bounded
with respect to $\widetilde{H}_m$. But, as in the proof of 
exponential decay (Lemma \ref{exp}), we have for each $i=1,2,3$
\begin{equation}
(G_{\lambda}^i \widetilde{\Phi}_m, (\widetilde{H}_m-E^V(m,1))G_{\lambda}^i 
\widetilde{\Phi}_m) = (\widetilde{\Phi}_m,
|\nabla_{x} G_{\lambda}^i|^2 \widetilde{\Phi}_m)
\end{equation}
and we arrive at the bound
\begin{equation}
\Sigma \kern-10pt \int |q(k,\lambda)|^2 \|a_{\lambda}(k)\widetilde{\Phi}_m\|^2 
{\rm d}k \leq C\Sigma \kern-10pt \int_{|k|
\leq \Lambda} \frac{|q(k,\lambda )|^2}{\omega(k)^2}  \Bigl[\Vert G_{\lambda} \widetilde{\Phi}_m
\Vert^2 +\Vert |\nabla_x G_{\lambda}| \widetilde{\Phi}_m \Vert^2
+\omega(k)^2 \|w_{\lambda} \widetilde{\Phi}_m\|^2 \Bigr] {\rm d}k  \
,\label{infbound1}
\end{equation}
where $C$ is some constant independent of $m$. Since $q(k,\lambda)$ is
arbitrary we obtain for almost every $k$ and each $\lambda$ that
\begin{equation}
 \|a_{\lambda}(k)\widetilde{\Phi}_m\|^2 \leq
C\omega(k)^{-2}  \left[\Vert G_{\lambda} \widetilde{\Phi}_m \Vert^2 + \Vert |\nabla_{x}
G_{\lambda}| \widetilde{\Phi}_m \Vert^2
+\omega(k)^2 \|w_{\lambda}\widetilde{\Phi}_m\|^2\right] \chi_{\Lambda}(k)  \ .
\end{equation}
The right side is bounded by
\begin{equation}
 \frac{C}{|k|} \Vert |x| \widetilde{\Phi}_m \Vert^2 \chi_{\Lambda}(k) \ , \label{infbound2}
\end{equation}
which is finite on account of the exponential decay of $\widetilde{\Phi}_m$.
\medskip
\end{proof}
\begin{proof}[Proof of Theorem \ref{finitederiv}]
First some notation: For
any function $f(k)$ define
\begin{equation}
(\Delta_{h} f)(k) = f(k+h)-f(k),
\end{equation}
and
\begin{equation}
\Delta_{-h}a(f) = a(\Delta_h f)
\end{equation}
Returning to (\ref{keyineq}) with $f$ replaced by $\Delta_h f$ we
have
\begin{eqnarray}
& & (\Delta_{-h}a(f) \widetilde{\Phi}_m, a(\omega \Delta_{h} f) \widetilde{\Phi}_m) 
\leq \\ \nonumber
& - & 2(\Delta_{-h}a(f)\widetilde{\Phi}_m, (\Delta_{h} f,G)
(p+\widetilde{A})\widetilde{\Phi}_m)
+  i(\Delta_{-h}a(f)\widetilde{\Phi}_m, (\Delta_{h}f,\omega w)
\widetilde{\Phi}_m)
\end{eqnarray}
which can be rewritten as
\begin{eqnarray}
& &(\Delta_{-h}a(f) \widetilde{\Phi}_m, \Delta_{-h}a( \omega f) 
\widetilde{\Phi}_m) \label{differenz}  \\ 
& \leq & (\Delta_{-h}a(f) \widetilde{\Phi}_m, a((\Delta_h \omega) f(\cdot +h))
\widetilde{\Phi}_m)  \nonumber \\
& - & 2(\Delta_{-h}a(f) \widetilde{\Phi}_m, (f,\Delta_{-h} G)  
(p+\widetilde{A}) \widetilde{\Phi}_m) 
 +  i(\Delta_{-h}a(f)\widetilde{\Phi}_m, (f,\Delta_{-h}(\omega w))
\widetilde{\Phi}_m) \ . \nonumber
\end{eqnarray}
Notice that without the first term on the right of the inequality sign, the
structure of this inequality is the same as (\ref{keyineq}),
except that, of course, $\Delta_{-h}a(f)$ plays the role of
$a(f)$, $\Delta_{-h} G$ plays the role of $G$
and $\Delta_{-h} (\omega w)$ plays the role of $\omega w$ . Thus,
without this term we would obtain immediately the estimate
analogous to (\ref{infbound1}),
\begin{eqnarray}
& & \Sigma \kern-10pt \int |q(k,\lambda)|^2\Vert (\Delta_{-h}a_{\lambda})(k) 
\widetilde{\Phi}_m \Vert^2  {\rm d}k  \nonumber \\
& \leq & C \Sigma \kern-10pt \int \frac{|q(k,\lambda)|^2}{\omega(k)^2} \Bigl[\Vert
\Delta_{-h}G_{\lambda} \widetilde{\Phi}_m \Vert^2 + \Vert |\nabla_x \Delta_{-h}G_{\lambda}|
\widetilde{\Phi}_m \Vert^2 + \omega^2 \Vert  
\Delta_{-h} (\omega w_{\lambda})\widetilde{\Phi}_m \Vert^2 \Bigr] {\rm d}k  \ . 
\label{infbound3}
\end{eqnarray}
The remaining term in equation (\ref{differenz}), after summing over the functions
$qg_i/\sqrt{\omega}$, turns into
\begin{equation}
\Sigma \kern-10pt \int \left( (\Delta_{-h}a_{\lambda}(k) \widetilde{\Phi}_m, a_{\lambda}
(k-h)\widetilde{\Phi}_m \right)
\frac{|q(k,\lambda)|^2}{\omega(k)} (\Delta_h \omega)(k-h){\rm d}k
\end{equation}
which, by Schwarz's inequality, is bounded above by
\begin{equation}
\left[ \Sigma \kern-10pt \int |q(k,\lambda)|^2 \Vert \Delta_{-h}a_{\lambda}(k) \widetilde{\Phi}_m\Vert^2 {\rm d}k
\right]^{1/2} \left[\Sigma \kern-10pt \int \| a_{\lambda}(k-h)\widetilde{\Phi}_m\|^2
\frac{|q(k,\lambda)|^2}{\omega(k)^2}|\Delta_h\omega(k-h)|^2 {\rm d}k
\right]^{1/2} \ .
\end{equation}
This, together with (\ref{infbound3}), yields
\begin{multline}
\Sigma \kern-10pt \int |q(k,\lambda)|^2\Vert (\Delta_{-h}a_{\lambda}(k) \widetilde{\Phi}_m \Vert^2  {\rm d}k \\
\leq C \Sigma \kern-10pt \int \frac{|q(k,\lambda)|^2}{\omega(k)^{2}}  \Bigl[\Vert
\Delta_{-h}G_{\lambda}
\widetilde{\Phi}_m \Vert^2 + \Vert |\nabla_x \Delta_{-h}G_{\lambda}| 
\widetilde{\Phi}_m \Vert^2 + \omega(k)^2 \Vert \Delta_{-h}(\omega w_{\lambda}) 
\widetilde{\Phi}_m \Vert^2\Bigr] {\rm d}k  \\
+ C\Sigma \kern-10pt \int \frac{|q(k,\lambda)|^2}{\omega(k)^2}\| a_{\lambda}(k-h)\widetilde{\Phi}_m\|^2
|\Delta_h\omega(k-h)|^2 {\rm d}k \ .
\end{multline}
Again, since $q$ is arbitrary we obtain for every fixed ${\lambda}$
\begin{multline}
\Vert (\Delta_{-h}a_{\lambda}(k) \widetilde{\Phi}_m \Vert^2 \leq \frac{C}{\omega(k)^2}
\Bigl[\Vert \Delta_{-h}G_{\lambda} \widetilde{\Phi}_m \Vert^2 +
 \Vert |\nabla_x \Delta_{-h}G_{\lambda}| \widetilde{\Phi}_m \Vert^2\\
+ \omega(k)^2 \Vert \Delta_{-h}(\omega w_{\lambda}) 
\widetilde{\Phi}_m \Vert ^2 + \| a_{\lambda}(k-h)\widetilde{\Phi}_m \|^2 |\Delta_h\omega(k-h)|^2\Bigr] \ .
\end{multline}
Combining this with (\ref{infbound2}) we get
\begin{multline}
\Vert (\Delta_{-h}a_{\lambda}(k) \widetilde{\Phi}_m \Vert^2 \leq \frac{C}{\omega(k)^2}
\Bigl[\Vert \Delta_{-h}G_{\lambda} \widetilde{\Phi}_m \Vert^2 +
 \Vert |\nabla_x \Delta_{-h}G_{\lambda}| \widetilde{\Phi}_m \Vert^2
 + \omega(k)^2 \Vert \Delta_{-h}(\omega w_{\lambda}) 
\widetilde{\Phi}_m \Vert^2 \Bigr]   \\
+ \frac{C}{\omega(k)^2|k-h|} \Vert |x| \widetilde{\Phi}_m \Vert^2
\chi_{\Lambda}(k-h) |\Delta_h\omega(k-h)|^2 \ .
\end{multline}
The polarization vectors defined in  
(\ref{eq:pol1}), (\ref{eq:pol2}),
are differentiable away from the 3-axis. The same
straightforward estimates as in Section~\ref{sec:infrared} lead to
\begin{equation}
\Vert (\Delta_{-h}a_{\lambda}(k) \widetilde{\Phi}_m \Vert^2 \leq
C\left[\frac{1}{|k|(k_1^2+k_2^2)}+ \frac{1}{|k-h|\left((k_1-h_1)^2+(k_2-h_2)^2
\right)} \right]|h|^2 \Vert
|x| \widetilde{\Phi}_m \Vert^2 \label{diffbound}
\end{equation}
which hold for all $|k|<\Lambda$ and small $|h|$ with a constant $C$ that is 
independent of $m$.

Next, we observe that for $k \not= 0$ fixed, there exist a sequence of $h$ values,
say $h_l$,
tending to zero so that $h^{-1}(\Delta_{-he_j}a_{\lambda}(k) \widetilde{\Phi}_m$ converges weakly to some
element $v_j(k)$ which satisfies the estimate
\begin{equation}
\Vert v_j(k) \Vert^2 \leq C\frac{1}{|k|(k_1^2+k_2^2)} \Vert |x| 
\widetilde{\Phi}_m \Vert^2 \ .
\end{equation}
Here $e_j$ is the $j$-th canonical basis vector.
Next we identify $v_j(k)$ as the weak derivative of $a_{\lambda}(k)\widetilde{\Phi}_m$.
This weak derivative, by definition, can be computed via the expression
\begin{equation}
-(\Psi, a(\partial_j \phi)\widetilde{\Phi}_m)
\end{equation}
where $\Psi$ is any state in ${\cal H}$ and $\phi$ is any test function 
in $C^{\infty}_c(\mathbb{R}^3)$.
Clearly the above expression equals
\begin{equation}
\lim_{h \to 0} \Sigma \kern-10pt \int (\Psi,\Delta_{-he_j}a_{\lambda}(k)\widetilde{\Phi}_m) 
\phi(k) {\rm d}k \ .
\end{equation}
But along the sequence $h_l$
\begin{equation}
\lim_{l \to \infty}\Sigma \kern-10pt \int (\Psi,\Delta_{-h_le_j}a_{\lambda}(k)
\widetilde{\Phi}_m) \phi(k) {\rm d}k
= \int (\Psi,v_j(k)) \phi(k) {\rm d}k \ ,
\end{equation}
which identifies $v_j(k)$ as the (negative) weak derivative of $a_{\lambda}(k) 
\widetilde{\Phi}_m$.
\end{proof}

marcel@math.uab.edu

lieb@princeton.edu

loss@math.gatech.edu
\end{document}